\documentstyle[preprint,eqsecnum,aps,epsfig,pre]{revtex}

\begin{document}
\renewcommand{\thefootnote}{\fnsymbol{footnote}}
\draft
\title{Magnetization of ferrofluids with dipolar interactions: A Born--Mayer
expansion}
\author{B.~Huke and M.~L\"{u}cke \\}
\address{Institut f\"{u}r Theoretische Physik,Universit\"{a}t des Saarlandes,
 D-66041 Saarbr\"{u}cken, Germany \\}

\renewcommand{\thefootnote}{\arabic{footnote}}
\setcounter{footnote}{0}
\date{\today}
\maketitle

% ABSTRACT

\begin{abstract}

For ferrofluids that are described by a system of hard spheres interacting via
dipolar forces we evaluate the magnetization as a function of the internal
magnetic field with a Born--Mayer technique and an expansion in the dipolar
coupling strength. Two different approximations are presented for
the magnetization considering different contributions to a series expansion in 
terms of the volume fraction of the particles and the dipolar coupling
strength.

\end{abstract}


% \begin {multicols}{2}
\narrowtext

% BEGIN OF THE TEXT

\section{Introduction}
\label{intro}

Ferrofluids \cite{R85} are suspensions of ferromagnetic particles of about 
10 nm diameter in a carrier fluid. The particles are stabilized against 
aggregation by coating with polymers or by electrostatic repulsion of charges 
brought on their surface. On macroscopic scales, ferrofluids can be 
described as liquids with intrinsic superparamagnetic properties.

In this paper, we are concerned with the equilibrium magnetization $M$ as a 
function of the internal magnetic field $H$ for given temperature and particle 
concentration. Sufficiently low concentrated ferrofluids behave like a 
paramagnetic gas. Therein the interaction between the particles can be 
neglected and the equilibrium magnetization can be described properly by the 
Langevin function. The magnetic properties are then necessarily weak.
To produce ferrofluids with strong magnetic properties one has to have 
either a higher particle concentration or one has to use ferromagnetic 
material with a large bulk magnetization, e.~g., cobalt instead of magnetite.
In both cases the  magnetization is strongly influenced
by dipole--dipole and other interactions between the particles. 

Several models of dipolar interacting systems have been studied in the 
literature.
Numerical investigations were based on density functional approaches
\cite{GD94,GD96,GD97a,GD97b,GD98} 
and Monte Carlo simulations \cite{LS93,WL93a,WL93b,SG95a,SG95b,SPLW,TWT99,CSP00}. 
The models differ in the treatment
of the short range interactions, which were described by hard sphere 
\cite{GD97a,WL93a,WL93b,TWT99,CSP00,S96}, 
other hard core potentials \cite{GD97a,SPLW}, 
soft sphere \cite{LS93,SG95a}, 
or Lenard--Jones potentials \cite{GD94,GD96,GD98,LS93,SG95b}. 
These investigations were mainly undertaken
to reveal the phase transition properties. These properties are substantially
different for different short range interactions.  
Thus, for example the question whether a system of particles interacting via
long range dipolar forces shows without any dispersive energy, e.~g., from
attractive van der Waals energy a "liquid--vapor" phase coexistence of a dense
and a less dense phase is currently being discussed 
\cite{LS93,SPLW,CSP00,S96,R96,TTO97,L99}.

Analytical models focus mainly on the equilibrium magnetization in the gas 
phase (were the term "gas" refers, as far as ferrofluids are addressed, to the 
magnetic particle subsystem within the liquid carrier). Such models are
the Onsager model \cite{Ons36}, the Weiss model \cite{GBCMPC83},
the mean spherical approximation \cite{W71}, and an approach by Buyevich and 
Ivanov \cite{BI92} (called high temperature approximation in \cite{P95}). 
These models were tested experimentally for ferrofluids 
\cite{P95,ML90,SPMS90,PML}. Especially the mean spherical model
and the high temperature approximation showed good results \cite{P95}. 

Our approach assumes the magnetic particles in the ferrofluid to be hard 
spheres with a common 
diameter $D$ and dipolar moment $m$. We use the technique of the 
Born--Mayer expansion \cite{HMcD90} together with
an expansion in the strength of the dipolar coupling to get analytical 
approximations. They are obtained via series expansions of the free
energy in terms of two parameters: (i) the volume fraction of the hard
core particles $\phi$ and (ii) a dimensionless dipolar coupling constant 
$\epsilon$, given by the ratio between a typical dipolar energy for particles 
in hard core contact and the thermal energy $kT$.
Our result for the magnetization goes beyond 
the high temperature approximation \cite{BI92} and reduces to it in 
linear order in $\phi$ and $\epsilon$.

Dipolar forces fall off as $r^{-3}$ and are thus of long range nature. 
This long range character requires great care when invoking the thermodynamic 
limit \cite{BGW98,WZ95,GD95}. To circumvent the problem we model the 
dipolar fields that are generated by distant particles by a magnetic continuum 
field (similar to the treatment in the Weiss model) while incorporating the 
near field contributions explicitly in a statistical mechanical description. 
The magnetization $M$ is then 
derived as a function of the internal magnetic field $H$. The so obtained
relation $M(H)$ is independent of the probe geometry. Once $M(H)$ is known, 
the magnetization for a given geometry can in principle be derived by solving 
the macroscopic Maxwell equations. This may practically still be a difficult 
task, at least as long as the external field is small or absent. In this 
case it is known  that the magnetization will show for general shaped probes a 
nontrivial spatial variation at high enough densities \cite{GD97b,GD98}.
 
Since our method yields an expression for the free energy of the model 
system we can in principle calculate also other thermodynamic quantities 
and in particular address the questions of phase transition,
e.~g., between gas and liquid or between ferromagnetic and non--ferromagnetic 
phases. We have not addressed the question of a gas--liquid transition of
the magnetic particles suspended in the ferrofluid since it is believed that 
short range van der Waals--like attractions would have to be incorporated to
model real ferrofluids appropriately in this regard \cite{SG95a}. 
However, the question whether a strong dipolar coupling induces in zero
external field a spontaneous magnetization that is currently debated in the
literature \cite{GD94,GD96,WL93a,WL93b,ZW93,ZW94} 
is briefly touched upon in Sec.~\ref{ferromag} of this paper.

The paper is organized as follows: In Sec.~\ref{model} we discuss the 
connections between the various fields that are of relevance in a ferrofluid.
We present the model to treat the long range dipolar forces. In Sec.~\ref{bm}
we present the expansion method to get analytical solutions in terms of 
the two small parameters $\epsilon$ and $\phi$. In Sec.~\ref{exp1} 
we calculate an expression for the magnetization that contains only linear
terms in $\phi$ but, at least in principle, arbitrary high orders in $\epsilon$.
In Sec.~\ref{exp2} a different expression is derived containing also
quadratic terms in $\phi$ but also only up to second order terms in $\epsilon$.
In Sec.~\ref{discuss} we discuss our findings and
investigate the applicability of the results in the $\phi$--$\epsilon$ plane. 
Sec.~\ref{conclude} contains a short conclusion.

\section{Magnetic fields and magnetization}\label{model}

We are interested in the effect of dipolar interactions of the magnetic
particles in a ferrofluid on the equilibrium magnetization of the ferrofluid. 
To that end we consider the ferrofluid as an ensemble of
identical spherical particles of diameter $D$, each carrying a magnetic moment
of magnitude $m$. These particles interact with each other via magnetic
dipole--dipole interactions and a hard core repulsion with hard core diameter 
$D$. We assume $D \geq D_{mag}$ where $D_{mag}$ is the diameter of the
magnetic core of the particles thus allowing for a surfactant surface layer that
provides a steric repulsion.

The particles can lower their potential energy by orienting their
magnetic moments parallel to a local magnetic field. However any interaction of
the particles with the fluid medium they are suspended in is ignored. The latter
is taken to be magnetically inert.

\subsection{Different magnetic fields}
\label{fields}

Before we outline in Sec.~\ref{decomp} how we determine in principle the
magnetization of the ferrofluid we should like to review briefly the
different magnetic fields that one has to distinguish and that are of importance
in a system with dipolar interactions. The first field is the 
external magnetic field ${\bf H}_e$ that is applied 
outside of the probe. If dipolar interactions can be neglected, 
${\bf H}_e$ is also the field at the position of the particles -- 
at least as long as the carrier fluid can be treated a magnetic vacuum which
we will assume throughout the paper. In the presence of dipolar interactions
additional fields have to be considered. One of them is the internal field 
${\bf H}$, that is the macroscopic field inside the probe. By assuming that the
equation of state ${\bf M}({\bf H})$ is known, ${\bf H}$ can be calculated
by the common methods of continuum magnetostatics. But the macroscopic field 
${\bf H}$ differs in general from the field ${\bf H}_{local}$ that the 
magnetic particles feel.

So far one has employed in the ferrofluid literature two models to calculate 
${\bf H}_{local}$ from ${\bf H}$ that are similar in spirit, namely the 
Weiss model \cite{GBCMPC83} and the Onsager model \cite{Ons36}. Both introduce a
virtual hollow sphere inside a magnetic continuum such that the sphere contains 
a single magnetic particle in its center. The Weiss 
model assumes the magnetization
${\bf M}$ and the internal field ${\bf H}$ to be constant everywhere in the
magnetic continuum surrounding the sphere. Then the field inside the sphere is 
given by
\begin{equation}
{\bf H}_s = {\bf H} + \frac{\bf M}{3} \;\; . \label{2HlocWeiss}
\end{equation}
This is the field that the single magnetic particle feels within the Weiss
model, i.~e.\ ${\bf H}_{local} = {\bf H}_s$.

The Onsager model, on the other hand, is restricted to linearly responding 
fluids and calculates the field inside the sphere on the assumption that it is
really hollow and that therefore ${\bf H}$ and ${\bf M}$
differ  near the sphere from its bulk values. In that case the field within the
sphere is
\begin{equation}
{\bf H}_{local} = \frac{3 \chi + 3}{2 \chi + 3} {\bf H} \;\; .
\label{2HlocOns}
\end{equation}
$\chi$ is the susceptibility.  
In both models the magnetization is calculated as the magnetization of a system
of {\em noninteracting\/} dipoles in the magnetic field ${\bf H}_{local}$, 
i.~e., 
\begin{equation}
M = M_{sat} {\cal L} \left( \frac{m}{kT} H_{local} \right) \label{2Meqn} \;\; .
\end{equation}
Here
\begin{equation}
M_{sat} = \frac{N}{V} \frac{m}{\mu_0}  \label{Msat} 
\end{equation}
is the saturation magnetization of the fluid, ${\cal L}$ 
the Langevin function, $m$ the magnetic moment of the particles, and $N/V$ their
number density.
In the Onsager model the Langevin function is consistently used only in linear 
order. Letting $M = \chi H$ on the left hand side of (\ref{2Meqn}) and using 
(\ref{2HlocOns}) allows to calculate $\chi$. 

In the Weiss model the 
selfconsistent solution $M(H)$ is determined using (\ref{2Meqn}) and  
(\ref{2HlocWeiss}).
The Onsager and Weiss models differ in the treatment of the back reaction of the 
particle inside the sphere on the magnetic continuum near the sphere's boundary.

\subsection{Decomposition of fields}
\label{decomp}

Since the magnetic continuum is a macroscopic concept one should be careful
when using it on the mesoscopic length scales of the interparticle distances 
and particle diameters. A first--principle statistical mechanical calculation 
of the magnetization would start with expressing the energy of the system in 
terms of the statistical variables of the constituents. In this context the
local magnetic field ${\bf H}_{local}$ that a magnetic moment, say, at position
${\bf x}_i$ feels is of
importance. It is composed of two different magnetic fields, the
external field ${\bf H}_e$ and the dipolar contribution ${\bf H}_{dipole}$
from the other $N-1$ particles at positions ${\bf x}_j$, possessing a magnetic 
moment ${\bf m}_j$. Thus 
within the first--principles approach one has 
${\bf H}_{local} = {\bf H}_e + {\bf H}_{dipole}$,
where
\begin{equation}
\label{Hdiptotal}
{\bf H}_{dipole}({\bf x_i}) = \sum_{j} 
\frac{ 3 {\bf \hat{r}}_{ij} ({\bf m}_j \cdot 
{\bf \hat{r}}_{ij} ) - {\bf m}_j}
{4 \pi \mu_0 r_{ij}^3}   \;\; .
\end{equation}  
Here ${\bf r}_{ij} = {\bf x_i} - {\bf x_j}$,  $r_{ij} = |{\bf r}_{ij}|$, and 
${\bf \hat{r}}_{ij} = {\bf r}_{ij}/r_{ij}$.

The long range character of the dipolar forces requires special care when
invoking the thermodynamic limit $V \rightarrow \infty$ and $N \rightarrow \infty$
($N/V =$ const.) --- for a critical discussion see, e.g., 
Ref.~\cite{WZ95,BGW98}.
The reason is that the dipolar contribution (\ref{Hdiptotal}) will in general 
depend
on the geometry of the ferrofluid probe and the location of ${\bf x_i}$ within
it. Thus the equilibrium magnetization of a probe in an external field will in
general depend on the geometry of the latter and furthermore it will be
spatially varying. We therefore use here an approach similar to the one
that has been used successfully in solid state theory \cite{AM76} to 
determine, e.~g., the crystal field splitting caused
by local fields. It properly accounts for the contributions from
microscopic and macroscopic scales. 

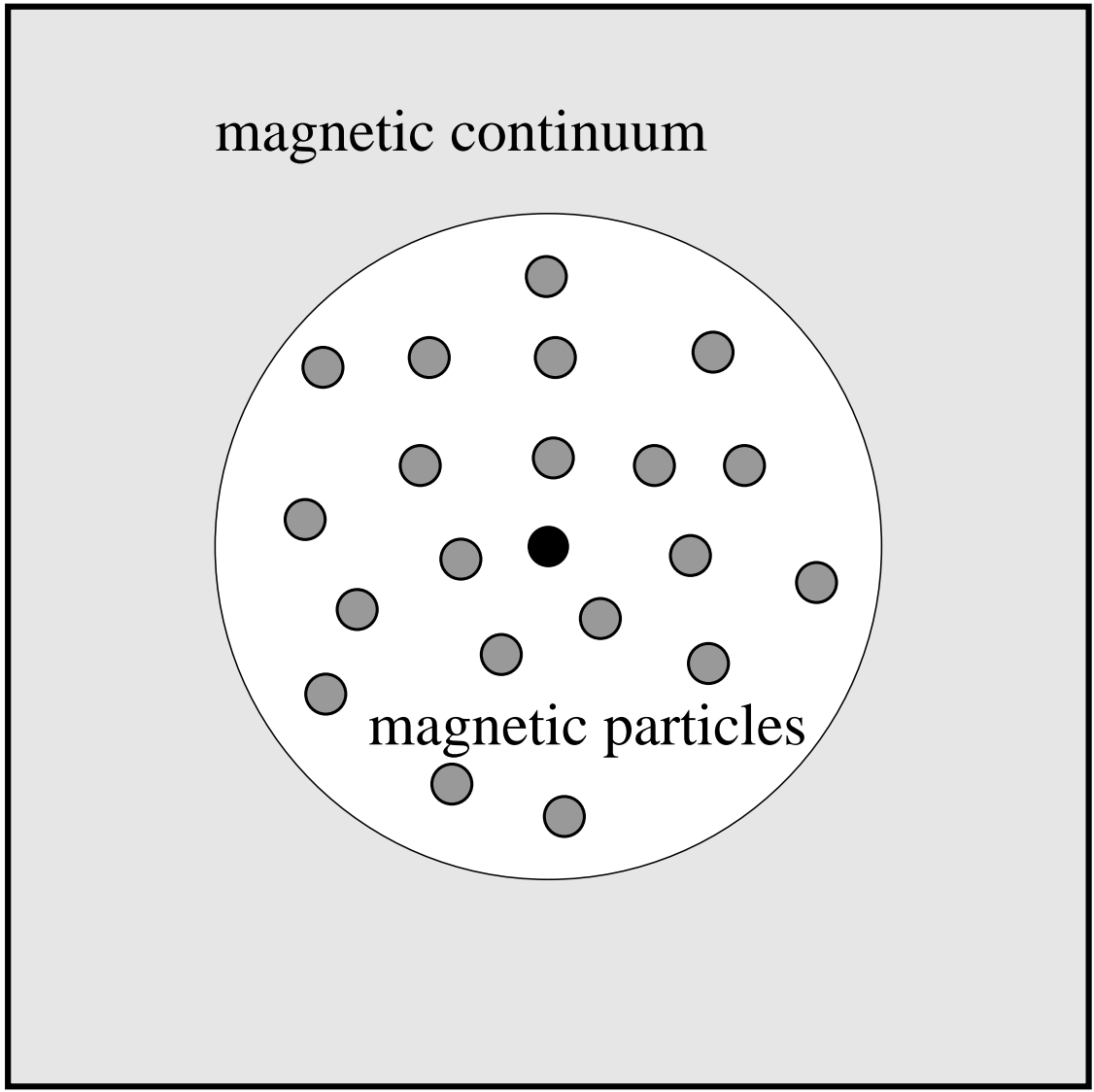
\begin{figure}\caption[]
{Geometry of our model. Every particle (black) feels the magnetic near field 
generated by the dipoles of the neighbours (dark gray) within the radius $R_s$ 
and a contribution from the continuum (light gray) that models the fields of 
the far--away particles and the external field.}
\label{ourmod}
\end{figure}

Consider some magnetic particle $i$ in a ferrofluid probe in thermodynamic
equilibrium. The particles beyond some distance 
$R_s$ from ${\bf x}_i$ can be considered as independent from particle $i$ if
$R_s$ is larger than the correlation length induced by the dipolar interactions. 
Furthermore, if $R_s$ is large enough, their contribution to the local field 
at ${\bf x}_i$ can be approximated by a contribution from a magnetic continuum 
with equilibrium magnetization ${\bf M}$ and macroscopic field ${\bf H}$. We 
assume that the distance $R_s$ is still small compared to the length scale on 
which the macroscopic fields ${\bf M}$ and ${\bf H}$ vary.
Thus we introduce a virtual sphere of radius $R_s$ around particle $i$
(dark particle in the center of Fig.~\ref{ourmod})
to separate the dipolar field into a "far" and a "near" contribution 
\begin{equation}
{\bf H}_{dipole,far}({\bf x}_i) = \sum_{r_{ij} > R_s} 
\frac{ 3 {\bf \hat{r}}_{ij} ({\bf m}_j \cdot 
{\bf \hat{r}}_{ij} ) - {\bf m}_j}
{4 \pi \mu_0 r_{ij}^3}   \;\; ,
\end{equation}  
\begin{equation}
\label{Hdipnear}
{\bf H}_{dipole,near}({\bf x}_i) = \sum_{r_{ij} < R_s} 
\frac{ 3 {\bf \hat{r}}_{ij} ({\bf m}_j \cdot 
{\bf \hat{r}}_{ij} ) - {\bf m}_j}
{4 \pi \mu_0 r_{ij}^3}   \;\; .
\end{equation}  
Then
\begin{equation}
{\bf H}_{local} = {\bf H}_e + {\bf H}_{dipole,far} + {\bf H}_{dipole,near} 
\;\; .
\end{equation}
If the sphere would be empty, ${\bf H}_e + {\bf H}_{dipole,far} = {\bf H}_s$ 
would be the local field inside the sphere. 

A key point of our treatment is to
express the far field ${\bf H}_{dipole,far}$ within the continuum approximation.
Using this approach the field in the empty sphere 
is given by ${\bf H}_s = {\bf H} + {\bf M}/3$ (\ref{2HlocWeiss}). Note that 
this approximation is not valid near the sphere's boundary. But at the center 
of the sphere ${\bf H}_{local}$ consists of the continuum  contribution 
${\bf H}_s$ from the far region and the contribution for the dipoles 
within the sphere: 
\begin{equation}
{\bf H}_{local} = {\bf H}_s + {\bf H}_{dipole,near} = 
{\bf H} + \frac{{\bf M}}{3} + {\bf H}_{dipole,near} \;\; .
\end{equation}
This result agrees with Eq.~(27.26) of Ref. \cite{AM76} where it has been 
derived with slightly different arguments for electric dipoles. 
Due to the long range character of the dipolar forces ${\bf H}_{dipole}$ will 
be in general geometry dependent and spatially varying. ${\bf H}_s$, 
respectively ${\bf H}$ and ${\bf M}$ will then also show these features as 
mentioned above.

If the dipolar coupling between the particles is so weak that even the 
dipolar fields of the nearest neighbours of particle $i$ can be described 
by a continuum field, i.~e., if $R_s$ can be chosen as being smaller than the
mean distance between the particles, we can drop the
contribution ${\bf H}_{dipole,near}$ altogether and arrive at the Weiss model,
where a single particle is located inside the hollow sphere in the continuum. 

\subsection{Equilibrium magnetization}

We want to determine the thermodynamic equilibrium relation $M(H)$ between the
magnetization
\begin{equation}
{\bf M} = \frac{1}{\mu_0 V} \sum_i \left< {\bf m}_i \right> =  
\frac{N}{\mu_0 V} \left< {\bf m} \right> \label{stamechprob}
\end{equation}
and the macroscopic magnetic field ${\bf H}$ in the thermodynamic limit.
Instead of considering the dipolar interaction of all particles in an 
external field in the statistical mechanical problem (\ref{stamechprob})
we take explicitly only interactions between those particles 
into account whose distance is smaller than the sphere radius $R_s$. The other 
interactions are represented by the far--field continuum approximation 
${\bf H}_s = {\bf H} + {\bf M}/3$. 
The magnetization $M_{sphere}(H_s)$ resulting from this decomposition of fields
is then identified with the
equilibrium magnetization $M(H)$ of the ferrofluid
\begin{equation}
M_{sphere}(H + M/3) = M(H) \;\; . \label{Ms_M_H_rela}
\end{equation}
Thus after having obtained the approximate expressing for $M_{sphere}$ as a
function of $H+M/3$ we then obtain from solving (\ref{Ms_M_H_rela}) for $M$ an
approximation for the sought after equilibrium relation $M(H)$. 
The functional dependence of $M$ on $H$ is independent of the probe geometry.

In the limit of weak dipolar coupling or when $R_s$ becomes smaller than the
mean distance between the particles we find $M_{sphere} = 
M_{sat} {\cal L} \left[ \frac{m}{kT} \left( H + M/3 \right) \right]$ so that
the Weiss model is recovered as discussed above. 

The magnetization $M_{sphere}$ in (\ref{Ms_M_H_rela}) depends on two
dimensionless parameters that characterize the thermodynamic state
of the ferrofluid. One of these parameters is the volume concentration of
the particles
\begin{equation}
\phi = \frac{N}{V} \frac{\pi D^3}{6} \label{phidef} \;\; .
\end{equation}
The ratio $\phi_{mag}$ of the volume of the magnetic material to the total 
volume is $\phi_{mag}=(D_{mag}/D)^3 \phi$. The other parameter is
\begin{equation}
\epsilon = \frac{m^2}{4 \pi \mu_0 k T D^3} \label{epsidef} \;\; ,
\end{equation}
the ratio between a typical energy of dipole--dipole interaction of particles in
contact (i.~e., at the distance of the hard core diameter $D$) and the 
thermal energy $k T$.  

\section{Canonical partition function}\label{bm}

We use the canonical ensemble average to evaluate $M_{sphere}$. 
Given a system of $N$ interacting particles with an interaction potential
$V_{ij}$ ($1 < i,j < N$) and external potential per 
particle $V_i$, the canonical partition function is given by
\begin{equation}
Z = \int e^{ -\sum_k v_k - \sum_{i<j} v_{ij}} \;
d \Gamma \;\; . \label{candistrigen}
\end{equation} 
Here $v_i = V_i/kT$, $v_{ij} = V_{ij}/kT$, and $d \Gamma$ means integration 
over the configuration space. In our case, 
a configuration is characterized by specifying the position vector 
${\bf x}_i$ and two angles for each magnetic particle. The two angles
define the direction of the magnetic moment ${\bf m}_i$. The modulus $m$ is 
assumed to be constant and the same for all particles.
Note that we ignore any translational and rotational degrees of freedom of the
particles that carry the magnetic moments, since they have no effect on
the magnetization. Only the locations of the moments,
i.~e., of the particles and the orientations of the moments are considered as
statistical variables.

In the first--principles statistical mechanical problem identified by a
superscript $0$, the external potential would be
the energy of a dipole in the external magnetic field
\begin{equation}
V^{0}_i = - {\bf m}_i \cdot {\bf H}_e \;\; .
\end{equation}
The interparticle potential is modelled by a dipole--dipole (DD) 
interaction plus hard core (HC) repulsion. Thus
\begin{equation}
V^{0}_{ij} = V^{0,DD}_{ij} + V^{HC}_{ij} \;\; ,
\end{equation}
\begin{equation}
V^{0,DD}_{ij} = 
- \frac{ 3 ({\bf m}_i \cdot {\bf \hat{r}}_{ij} )({\bf m}_j \cdot 
{\bf \hat{r}}_{ij} ) - {\bf m}_i \cdot {\bf m}_j }
{4 \pi \mu_0 r_{ij}^3} \;\; ,
\end{equation}
\begin{equation}
V^{HC}_{ij} = \left\{ \begin{array}{cl} 
0 & \qquad \mbox{for $r_{ij} > D$} \\
\infty & \qquad \mbox{for $r_{ij} < D$} 
\end{array} \right. \;\; . 
\end{equation}

The replacement of the dipolar magnetic fields from far--away particles by a 
field that has its origin in a magnetic continuum results in a new canonical 
partition function with the "external" potential of a dipole
\begin{equation}
V_i = - {\bf m}_i \cdot {\bf H}_s \;\; , \label{Viconti}
\end{equation}
in the field ${\bf H}_s = {\bf H} + {\bf M}/3$ and a dipolar 
interaction term with a cutoff, i.~e.,
\begin{equation}
V^{DD}_{ij} = \left\{ \begin{array}{cl} 
V^{0,DD}_{ij} & \qquad \mbox{for $r_{ij} < R_s$} \\
0 & \qquad \mbox{for $r_{ij} > R_s$} 
\end{array} \right. \;\; . \label{Vijconti}
\end{equation}

Note that when using (\ref{Viconti}) and (\ref{Vijconti}) in the expression 
(\ref{candistrigen}) for the partition function we describe every particle 
$i$ as being at the center of a sphere of radius $R_s$ inside a magnetic
continuum such that each particle feels
the "external" field ${\bf H}_s$ and explicit dipolar fields 
${\bf H}_{dipole,near}({\bf x}_i)$ (\ref{Hdipnear}) from the 
particles whose distance is smaller than $R_s$.  

The aforementioned statistical mechanical problems with the long
range nature of the bare dipolar interactions is thus circumvented by the 
cutoff at $R_s$ in (\ref{Vijconti}) that results from decomposing dipolar 
fields into a near and a far contribution. Dipolar forces appear explicitly 
only as forces with a finite range. Their influence on the magnetization in our
approach is therefore independent of the geometry of the probe. The  
geometry dependence enters only via the effective "external" field $H_s$ from 
the far field contribution.

\subsection{Born--Mayer expansion method}

Since an integral such as (\ref{candistrigen}) is hard to solve even 
numerically we use the Born--Mayer expansion method \cite{HMcD90} to get 
analytical results. The key point of this method is to write
\begin{equation}
Z = \int \prod_k e^{-v_k}
\prod_{i<j} (1 + f_{ij} ) \; d \Gamma \;\; ,
\end{equation} 
where
\begin{equation}
f_{ij} = e^{- v_{ij}} -1 \;\; .
\end{equation} 
If the typical interaction energy is small compared to $kT$, the $f_{ij}$ can be
considered as small parameters, and $Z$ can be expanded into a series:
\begin{eqnarray}
Z &=& \int \prod_m e^{-v_m} \; d \Gamma + 
\nonumber \\
&& \int \prod_m e^{-v_m} \sum_{i<j} f_{ij} \; d \Gamma + \nonumber \\
&& \int \prod_m e^{-v_m}
\sum_{i<j} f_{ij} \sum_{k<l} f_{kl}\; d \Gamma + \nonumber \\
&& ... \;\; . \label{generalbmexp}
\end{eqnarray} 
These integrals can be factorized and are therefore easier to handle.

The first order of (\ref{generalbmexp}) contains terms like
\begin{equation}
\int \prod_m e^{-v_m}
e^{-v^{DD}_{12} - v^{HC}_{12}}
\; d\stackrel{\rightarrow}{\bf x} d\stackrel{\rightarrow}{\Omega} \;\; . 
\end{equation} 
Here $d\stackrel{\rightarrow}{\bf x} d\stackrel{\rightarrow}{\Omega}$ is an
abbreviation for $d{\bf x}_1 ... d{\bf x}_N d\Omega_1 .. d\Omega_N$, and 
$d\Omega_i$ means the integration over the possible orientations of 
${\bf m}_i$. 

\subsection{Expansion in powers of $v^{DD}$}
\label{epsexpansion}

Obviously even the first order still cannot be 
evaluated analytically. Therefore a second series expansion is made
\begin{mathletters}
\begin{equation}
f_{ij} = e^{-v^{HC}_{ij}} e^{-v^{DD}_{ij}} -1 
= f_{ij}^{(0)} + f_{ij}^{(1)} + f_{ij}^{(2)} + ... ,
\end{equation}
\begin{eqnarray}  
f_{ij}^{(0)} &=& \left( e^{-v^{HC}_{ij}} - 1 \right) \\
f_{ij}^{(\alpha)} &=& \frac{\left(-v^{DD}_{ij}\right)^\alpha}{\alpha!} 
e^{-v^{HC}_{ij}} 
... \;\; ,  \alpha \geq 1 \;\;.
\end{eqnarray}
\end{mathletters}
So we expand $f_{ij}$ in powers of the reduced dipolar interaction 
$v^{DD}_{ij}$. The integrals in (\ref{generalbmexp}) that remain to be solved 
are of the form
\begin{equation}
\int \prod_m e^{-v_m} 
f_{ij}^{(\alpha)} f_{kl}^{(\beta)} ... 
\; d\stackrel{\rightarrow}{\bf x} d\stackrel{\rightarrow}{\Omega} \;\; . 
\end{equation} 

We now introduce a modification of the common graphical representation of
Born--Mayer integrals as follows 
\begin{enumerate}
\item Every distinct particle that appears via 
interaction terms of the form $f_{ij}^{(\alpha)}$ is represented by a circle. 
\item A zeroth order interaction term $f_{ij}^{(0)}$ is represented by an 
overlap of the circles $i$ and $j$.   
\item First, second, ... order interaction is represented by one, two, ... lines
connecting the circles.
\end{enumerate}
Note that the representation of the zeroth order dipolar interaction 
by two overlapping circles is a reminder
that in this case the integrand is nonzero only if the particles
are assumed to be in a configuration in which they would indeed overlap.  

It turns out, that the expansion in terms of the $f_{ij}^{(\alpha)}$ means an 
expansion in powers of the two parameters $\epsilon$ and $\phi$, that define 
the thermodynamical system. Every line in a representing graph, i.~e., every
power of $v^{DD}_{ij}$ results in a factor $\epsilon$. Every $n$--particle 
subgraph in which all circles are connected to each other directly or 
indirectly gives a factor of $\phi^{n-1}$. In the next two sections we will
present two expansions considering different terms.
  
\section{Expansion up to first order in  $\phi$}\label{exp1}

In this section only terms up to $O(\phi)$ will be taken into account. 

\subsection{Partition function}

In $O(\phi)$ the canonical partition function reads
\begin{eqnarray}
Z &=& \int \prod_k e^{-v_k} 
\; d\stackrel{\rightarrow}{\bf x} d\stackrel{\rightarrow}{\Omega} + 
\nonumber \\
&& \int \prod_k e^{-v_k} \sum_{i<j} f_{ij} 
\; d\stackrel{\rightarrow}{\bf x} d\stackrel{\rightarrow}{\Omega} 
+ O(\phi^2) \;\; .
\end{eqnarray} 
The $f_{ij}$ have yet to be expanded in powers of $v_{ij}^{DD}$. 
Figure~\ref{clusterexp1} shows the corresponding graphs.
 There are $N(N-1)/2 \approx N^2/2$ ways to choose $i$ and $j$. Because all 
particles are identical one can write
\begin{eqnarray}
Z &=& \int \prod_k e^{- v_k} 
\; d\stackrel{\rightarrow}{\bf x} d\stackrel{\rightarrow}{\Omega} +
\nonumber \\
&& \frac{N^2}{2} \int \prod_k e^{- v_k} 
f_{12} \; d\stackrel{\rightarrow}{\bf x} d\stackrel{\rightarrow}{\Omega} 
+ O(\phi^2) \;\; .
\end{eqnarray} 
Integrating over most degrees of freedom results in
\begin{eqnarray}
Z &=& Z_0 + 
\frac{N^2}{2} z_0^{N-2} 
\int e^{ - v_1 - v_2} 
f_{12} \; d{\bf x}_1  d{\bf x}_2 d\Omega_1  d\Omega_2 \nonumber \\
&& + O(\phi^2) \label{Zinexp1} \;\; .
\end{eqnarray} 
Here
\begin{equation}
Z_0 = z_0^N \;\; ; \qquad z_0 = 4 \pi V \frac{\sinh \alpha_s}{\alpha_s}
\label{Z0def}
\end{equation}
is the partition function of a paramagnetic gas of noninteracting particles in
the field $H_s$ defining the Langevin parameter 
\begin{equation}
\alpha_s = \frac{m H_s}{kT} \;\; . 
\end{equation}

\subsection{Expansion in the dipolar interaction}

Now we expand $f_{12}$ appearing in (\ref{Zinexp1}) in a power series in
$\epsilon$. The $n$--th summand of this series contains integrals of the
form
\begin{equation}
A_n = \int e^{ - v_1 - v_2}  
f^{(n)}_{12} \; d{\bf x}_1  d{\bf x}_2 d\Omega_1  d\Omega_2 \;\; .
\label{Adef}
\end{equation}
$A_0$ is special. Here one gets
\begin{eqnarray}
A_0 &=& \left( 4 \pi \frac{\sinh \alpha_s}{\alpha_s} \right)^2
\int \left( e^{-v^{HC}_{12}} -1 \right) \; d{\bf x}_1  d{\bf x}_2
\label{A0calc} \\
&=& \frac{1}{V} z_0^2
\int \left( e^{-v^{HC}_{12}} -1 \right) \; d{\bf r}_{12} \nonumber \;\; .
\end{eqnarray}
The integrand vanishes if $r_{12} > D$. Otherwise its value is $-1$. Thus
\begin{equation}
A_0 = - \frac{4}{3} \pi \frac{D^3}{V} z_0^2 \;\; ,
\end{equation}
or by expressing the result in terms of $\phi$
\begin{equation}
A_0 = - \frac{8}{N} \phi z_0^2 \;\; . \label{A0result}
\end{equation}
For $n \geq 1$ we have
\begin{eqnarray}
A_n &=& \frac{1}{n!} \int e^{- v_1 - v_2} \nonumber \\
&& \times \left( - v_{12}^{DD} \right)^n e^{-v^{HC}_{12}}
d{\bf x}_1  d{\bf x}_2 d\Omega_1  d\Omega_2 \;\; . 
\end{eqnarray}
Switching from ${\bf r}_2$ to the relative coordinate ${\bf r}_{12}$ and
integrating over ${\bf r}_1$ gives a factor of $V$. Then ${\bf r}_{12}$ runs over 
the sphere volume. We introduce spherical coordinates, i.~e.,
\begin{eqnarray}
{\bf m}_1 &=& m ( \cos \varphi_1 \sin \vartheta_1,\sin \varphi_1 
\sin \vartheta_1, \cos \vartheta_1) \nonumber \\
{\bf m}_2 &=& m ( \cos \varphi_2 \sin \vartheta_2,\sin \varphi_2 
\sin \vartheta_2, \cos \vartheta_2) \\
{\bf r}_{12} &=& r_{12} ( \cos \varphi \sin \vartheta,\sin \varphi 
\sin \vartheta, \cos \vartheta) \nonumber \;\; .
\end{eqnarray}
The direction of the magnetic field defines the $z$--axis.
Then, the integral assumes the form
\begin{eqnarray}
A_n &=& \frac{V}{n!} \int 
e^{\alpha_s \cos \vartheta_1 + \alpha_s \cos \vartheta_2} \nonumber \\
&& \times \left( \frac{m^2}{4 \pi \mu_0 k T r^3_{12}} \right)^n 
P^n(\varphi_1,\vartheta_1,\varphi_2,\vartheta_2,\varphi,\vartheta)\\
&& \times e^{-v^{HC}_{12}}
r^2_{12} dr_{12}  d\omega_{12} d\Omega_1  d\Omega_2 \;\; . \nonumber
\end{eqnarray}
The new spherical angle $\omega_{12}$ represents $\varphi$ and $\vartheta$. The
exact form of the function $P$ is not important, but $P$ and therefore $P^n$ is
a polynomial in the $\cos$ and $\sin$ of the six angles. Integration over
four of them can be done analytically. Finally this can also be done for 
$\vartheta_1$ and $\vartheta_2$ by substituting
$u_{1,2} = \cos \vartheta_{1,2}$. One gets an expression of the form
\begin{equation}
A_n = \frac{V}{n!} G^*_n(\alpha_s) \int_D^{R_s} 
\left( \frac{m^2}{4 \pi \mu_0 k T r^3_{12}} \right)^n 
r^2_{12} dr_{12} \;\; . \label{lastintegral}
\end{equation}
Here we have introduced the correct bounds of the last remaining integral
explicitly. By setting the lower bound to $D$ we have incorporated the hard 
core factor. The upper bound is given by the cutoff radius $R_s$
for the near-field dipolar contribution. While the evaluation of $G^*_n$ 
can be done analytically, it is quite difficult 
to do this by hand even for $n=2$. We therefore used
the computer algebra system {\em mathematica\/} to perform the integrations.  
See Appendix \ref{appa} for the form of the $G^*_n$.

For $n \geq 2$ one can safely set $R_s = \infty$ (see below). 
For $n=1$ this would result in a logarithmic divergence of the integral. But
$G^*_1 \equiv 0$ anyway, because the calculation of $G^*_1$ involves an 
averaging over a dipolar field. So by using $\epsilon$ and $\phi$ one finally
has
\begin{equation}
A_n = \frac{2 V^2}{N \pi (n-1) n!} G^*_n(\alpha_s) \epsilon^n \phi 
\qquad n \geq 2 \;\; ,
\label{Aninf}
\end{equation}    
\begin{equation}
A_1 = 0 \;\; .\label{A1result}
\end{equation}
Note that $A_1$ vanishes only in our spherical configuration with finite $R_s$. 
The divergence
of $A_1$ in the general, spatially unrestricted case is just an expression of 
fact that the dipolar forces are long range.
By treating the distant parts of the ferrofluid as a
continuum we incorporate any long-range effects and the resulting geometry 
dependence via the field $H_s = H + M/3$. Into this field enters the relation
between the external and the macroscopic internal field.

Two further comments should be made here. A generalization of our calculation 
for central symmetric interactions other than a hard sphere potential 
is possible. It requires an analytical or numerical evaluation of 
integrals of the form $\int r^{2 - 3 n}  e^{-v^{SR}} dr$
in (\ref{lastintegral}), with $v^{SR} = V^{SR}/kT$ and $V^{SR}$ denoting the 
$r$--dependent short range potential in question. 

The second thing is that we can now make quantitative 
statements about how large the virtual sphere has to be chosen. 
To ensure $A_1$ to vanish unambigously in (\ref{lastintegral}) and to introduce
$H_s$ instead of $H_e$ as "external" field $R_s$ has only to be finite.
The larger $R_s$ the better is the modeling of large--distance 
particle correlations entering into $A_n$ for $n>1$. Taking the limit $R_s
\rightarrow \infty$ as the final step in the calculation of the $A_n$ is
therefore appropriate from this point of view. On the other hand, the 
requirement of uniformity of the fields ${\bf H}$ and ${\bf M}$, that allows 
us to write $H_s = H + M/3$  restricts $R_s$ to values below the scale on 
which $H$ and $M$ vary. If one would use a  
finite radius $R_s$ one would get instead of (\ref{Aninf}) for $n \geq 2$
\begin{equation}
A_n = \frac{2 V^2}{N \pi (n-1) n!} G^*_n(\alpha_s) \epsilon^n \phi 
\left[1 - (D/R_s)^{3n-3} \right]  \label{AnwithRs}
\end{equation}
which allows an error estimate:
Consider a system where ${\bf M}$ and ${\bf H}$ do not vary on the scale of,
say, $\mu$m. For ferrofluids, $D \approx 10$ nm. Choosing 
$R_s = 10 D$ or $100 D$ is then both allowed and implies a difference in $A_2$
of about 0.1 percent.
The result for $R_s = 100 D$ is better than for $R_s = 10 D$, because in the
latter case particles in a distance range between 100 nm and 1 $\mu$m are 
treated in the continuum approximation and not correctly. But the error that 
is made by treating the ferrofluid as a
continuum already beyond $R_s = 10 D$ is only about 0.1 percent. We 
can safely assume that the macroscopic, magnetic properties do not vary on this 
scale. Thus 100 nm is an appropriate medium scale on which both requirements 
hold: The continuum approximation works well beyond this cutoff radius 
{\em and} the macroscopic fields ${\bf H}$ and ${\bf M}$ should be constant on 
this scale. Except for the calculation of $A_1$, it is then possible to set 
$R_s = \infty$ in the calculations of the integrals. 
   
Using the results (\ref{Aninf}), (\ref{A1result}), (\ref{A0result}), 
and (\ref{Adef}) in (\ref{Zinexp1}), 
one gets the following expression for $Z$:
\begin{equation}
Z = Z_0 \left[ 1 - 4 N \phi  + N \phi 
\sum_{n=2}^\infty  G_n(\alpha_s) \epsilon^n \right] 
 + O(\phi^2) \label{ZinOphi} \;\; .
\end{equation} 
Here we introduced the functions
\begin{equation}
G_n(\alpha_s) = \frac{1}{16 \pi^3 (n-1) n!} 
\left( \frac{\alpha_s}{\sinh \alpha_s} \right)^2 G^*_n(\alpha_s) \;\; , 
\label{Gdef}
\end{equation} 
some of which are given in Appendix \ref{appa}.

\subsection{Free energy and magnetization}

The next step is to compute the free energy
\begin{eqnarray}
\frac{F}{kT} &=& -\ln Z = 
-N \ln z_0
\nonumber \\
&& - \ln \left[ 1 - 4 N \phi  + N \phi 
\sum_{n=2}^\infty  G_n(\alpha_s) \epsilon^n \right] \\
&& + O(\phi^2) \nonumber \;\; .
\end{eqnarray} 
In $O(\phi)$, we can use $\ln(1 + x) = 1 + x$ here:
\begin{eqnarray}
\frac{F}{kT} &=& -N \ln z_0 
 + 4 N \phi  - N \phi 
\sum_{n=2}^\infty  G_n(\alpha_s) \epsilon^n \nonumber \\
&& + O(\phi^2) \;\; .  
\end{eqnarray} 
The magnetization turns out to be
\begin{eqnarray} 
\label{Msphere}
M_{sphere} (\alpha_s) &=& 
-\frac{1}{\mu_0 V} \frac{\partial F}{\partial H_s} = -\frac{m}{\mu_0 VkT}
\frac{\partial F}{\partial \alpha_s} \nonumber \\
&=& \frac{N m}{\mu_0 V} \left[ {\cal L}(\alpha_s) +  
\phi \sum_{n=2}^\infty  G'_n(\alpha_s) \epsilon^n \right] \\
&& + O(\phi^2) \;\; . \nonumber
\end{eqnarray}
The leading term is the Langevin function ${\cal L}$ times the
saturation magnetization $M_{sat} = \frac{N m}{\mu_0 V}$ of the fluid.

In order to determine $M(H)$ we identify, according to (\ref{Ms_M_H_rela}), 
$M_{sphere}(\alpha_s)$ with $M(\alpha)$, i.~e.,
\begin{eqnarray} 
\frac{M}{M_{sat}} &=& {\cal L}
\left(\alpha + \frac{mM}{3kT} \right) \label{Mimplicit} \nonumber\\
&& +  \phi \sum_{n=2}^\infty  G'_n \left(\alpha + \frac{mM}{3kT} \right) 
\epsilon^n + 
O(\phi^2) \;\; ,
\end{eqnarray}
where $\alpha$ is the usual Langevin parameter, 
\begin{equation}
\alpha = \frac{mH}{kT} \;\; .
\end{equation}
Instead of trying to find the function $M(\alpha)$ that solves this equation
exactly we expand the functions ${\cal L}$ and $G'_n$ for small $\phi$ into a 
series around $M=0$ and reinsert it on the right hand side. Using the fact that
$\frac{m M_{sat}}{3kT} = 8 \phi \epsilon$ grows linearly in $\phi$ we arrive at
\begin{mathletters} 
\label{MaginOphi1}
\begin{equation}
\label{Mexplicit} 
\frac{M(\alpha)}{M_{sat}} = L_{0,0} +  \phi \sum_{n=1}^\infty L_{1,n} \epsilon^n
+ O(\phi^2) 
\end{equation}
with 
\begin{eqnarray}
L_{0,0} &=& {\cal L}(\alpha) \label{L00} \\
L_{1,1} &=& 8{\cal L}(\alpha) {\cal L}'(\alpha) \label{L11} \\
L_{1,n} &=& G'_n(\alpha) \qquad \text{for} \qquad n \geq 2  \; . \label{L1n} 
\end{eqnarray}
\end{mathletters}
This is a consistent approximation in terms of $\phi$. On the other hand,
solving Eq.~(\ref{Mimplicit}) in a formally exact manner for $M$ would introduce 
higher orders in $\phi$ which we already neglected 
to arrive at Eq.~(\ref{Mimplicit}).

Note also that $M_{sphere} (\alpha_s)/M_{sat}$ (\ref{Msphere}) contains {\em
explicitly} a term $\sim \phi\epsilon^2$ as lowest nontrivial power coming from
the expansion in the near-field dipolar coupling strength. On the other hand the
self-consistent solution (\ref{MaginOphi1}) that solves Eq.~(\ref{Mimplicit})
starts out with a contribution $\sim \phi\epsilon$. The latter arises from the
far-field dipolar continuum via the magnetization $M$ in the dipole-induced
shift of the argument, $\alpha + \frac{mM}{3kT}$, of the Langevin function in 
Eq.~(\ref{Mimplicit}) --- in the absence of any dipolar interactions in the
system one would have $H_s = H_e = H$ leading to ideal paramagnetism.

\subsection{Comparison with previous results}

The Onsager model, the
Weiss model, and our calculation agree that up to order $\phi \epsilon$
\begin{equation} 
\frac{M}{M_{sat}} = {\cal L}(\alpha) 
+ 8 \phi \epsilon {\cal L}(\alpha) {\cal L}'(\alpha) + 
O(\phi^2) + O(\epsilon^2) \;\; . 
\end{equation}
This expression was also derived by Buyevich and Ivanov \cite{BI92} with 
a calculation similar to ours. However, they did not introduce 
a magnetic continuum approximation. Instead, they assumed a special probe 
geometry of a long cylinder parallel to the external magnetic field and 
performed an integration over all the particle's dipolar fields in the 
cylinder explicitly. The magnetization was therefore given in
terms of the external field. Their result agrees with ours because for the
cylindrical geometry chosen in \cite{BI92} ${\bf H}_e$ equals ${\bf H}$. 

A second paper that deals with our problem in a similar way was published by
Kalikmanov \cite{K92}. In
section 4, the author arrives at an equation for the magnetization that reads
in our notation
\begin{equation} 
\frac{M}{M_{sat}} = {\cal L}(\alpha) 
+ 3 \phi \epsilon^2 G'_2(\alpha) \int_1^\infty \frac{g_0(x)}{x^4} dx  \;\; . 
\label{Kalik}
\end{equation}
Here $g_0(x)$ is the hard sphere correlation function. In our
$O(\phi)$--approximation this function has to be set to one.
Then the $\phi \epsilon^2$--term agrees with ours. Note, however, that the 
above result (\ref{Kalik}) of Kalikmanov does not contain the 
$\phi \epsilon$--term resulting from the magnetic field from the continuum.
   
\section{Expansion up to second order in $\phi$ and $\epsilon$}\label{exp2}

It is possible to calculate $O(\phi^2)$--terms of the Born--Mayer 
expansion when $\epsilon$ is taken into account up to second order only. 
A more elegant way to calculate the magnetization in this approximation makes
use of the grand canonical rather than the canonical ensemble.
This approach allows to avoid the determination of some terms that
can be factorized into already known integrals and cancel out in the 
calculation of the free energy. However, the grand canonical approach has the 
disadvantage that it yields the magnetization as 
function of the chemical potential $\mu$ rather than the particle number $N$. 
Some more algebra is then required to find out the function $\mu(N)$. Here
we continue to work with the canonical ensemble.

\subsection{The graphs}

Figure~\ref{clusterexp2} shows the 12 additional graphs that are of second 
order in $\phi$ and of less than third order in $\epsilon$. Four of them
vanish because they contain at least one first--order dipolar interaction term
between otherwise unrelated particles. Integration over the relative position 
of these particles while leaving the relative positions between all other
particles and the direction of the magnetic moments fixed yields zero since it
involves a spatial averaging over a dipolar field. The graph labelled with the 
letter F vanishes for similar reasons that are explained in Appendix 
\ref{appb} where we calculate the integrals one by one. Their respective
contribution to the partition function is
\begin{mathletters}
\begin{eqnarray}
Z_A/Z_0 &=& 32 N \phi^2 \\
Z_B/Z_0 &=& -16 N \phi^2 \epsilon^2 G_2(\alpha_s)  \\
Z_C/Z_0 &=& 8 ( N^2 - 6N) \phi^2 \\
Z_D/Z_0 &=& -4(N^2-6N) \phi^2 \epsilon^2 G_2(\alpha_s) \\
Z_E/Z_0 &=& -5 N \phi^2 \\
Z_F/Z_0 &=& 0 \\
Z_G/Z_0 &=& \frac{1 + 6 \ln 2}{4} N \phi^2 \epsilon^2 G_2 (\alpha_s)\\
Z_H/Z_0 &=& - N \phi^2 \epsilon^2 K(\alpha_s) \;\; .
\end{eqnarray}
\end{mathletters}
The functions $G_2$ and $K$ are given in Appendix \ref{appa}.

\subsection{Free energy and magnetization}

Now we have all necessary terms at hand to calculate the canonical partition
function up to the desired order:
\begin{eqnarray}
&& \frac{Z}{Z_0} = 
1 - 4 (N-1) \phi +  (N-1) \phi \epsilon^2 G_2(\alpha_s) \nonumber \\
&& \qquad + 32 N \phi^2 -
16 N \phi^2 \epsilon^2 G_2(\alpha_s) + 8 (N^2 - 6 N) \phi^2 \nonumber \\
&& \qquad -4 (N^2 -6 N) \phi^2 \epsilon^2 G_2(\alpha_s) -5 N \phi^2 \\
&& \qquad +\frac{1+6 \ln 2}{4} N  \phi^2 \epsilon^2 G_2(\alpha_s) - 
N \phi^2 \epsilon^2 K(\alpha_s) + H.O.T. \;\; . \nonumber
\end{eqnarray}
The terms in $O(\phi)$ appear already in (\ref{ZinOphi}). They are presented 
here including the next higher order in $N$. The other terms come from 
$Z_A$ -- $Z_H$. 
To include all terms of $O(\phi^2, \epsilon^2)$ 
in the free energy one has to approximate the logarithm $\ln(1+x)$ 
by $x - x^2/2$. The quadratic 
order is necessary only for the $O(\phi)$--terms. New terms of $O(N^2)$ appear
and cancel against those of the terms from $Z_C$ and $Z_D$. 
One gets 
\begin{eqnarray}
&& \frac{F}{kT} = -N \ln z_0 
+ 4 N \phi + 5 N \phi^2 \nonumber \\
&& \qquad  - N \phi \epsilon^2 G_2(\alpha_s)
-\frac{1+6 \ln 2}{4} N  \phi^2 \epsilon^2 G_2(\alpha_s)\\
&& \qquad  + N \phi^2 \epsilon^2 K(\alpha_s) + H.O.T. \;\; . \nonumber 
\end{eqnarray}
The result is proportional to $N$ as it has to be.

The magnetization of the sphere is
\begin{eqnarray}
&& \frac{M_{sphere}(\alpha_s)}{M_{sat}} = {\cal L}(\alpha_s) + 
\phi \epsilon^2 G^\prime_2(\alpha_s) \label{protoM} \\
&& \qquad 
+\frac{1+6 \ln 2}{4} \phi^2 \epsilon^2 G^\prime_2(\alpha_s) 
- \phi^2 \epsilon^2 K^\prime(\alpha_s) + H.O.T. \;\; . \nonumber 
\end{eqnarray}
To calculate the magnetization as a function of $\alpha$ 
we identify (\ref{protoM}) with $M$ and use
again $\alpha_s = \alpha + \frac{mM}{3kT}$. The right hand side of (\ref{protoM})
has now to be expanded around $\alpha$ up to second order and the resulting 
equation has to be iterated twice to take into account all important terms
up to $\epsilon^2 \phi^2$. The result is
\begin{mathletters}
\label{MaginOphi2}
\begin{equation}
\frac{M(\alpha)}{M_{sat}} = L_{0,0} + \phi \epsilon L_{1,1} + 
\phi \epsilon^2 L_{1,2} +  \phi^2 \epsilon^2 L_{2,2} + ...
\end{equation}
with $L_{0,0}, L_{1,1}$, and $L_{1,2}$ defined in 
eqs.~(\ref{L00}) - (\ref{L1n}) and
\begin{eqnarray}
L_{2,2} &=& 64 {\cal L}(\alpha) {\cal L}^{\prime}(\alpha)^2 + 
32 {\cal L}(\alpha) {\cal L}^{\prime \prime}(\alpha) \nonumber \\
&& + \frac{1+6 \ln 2}{4} G^\prime_2(\alpha) - K^\prime(\alpha) \; .
\end{eqnarray}
\end{mathletters}

For the discussion in the next Sec. we decompose 
\begin{mathletters}
\label{L22decomp}
\begin{equation}
L_{2,2}(\alpha) = L_{2,2}^{sphere}(\alpha) + L_{2,2}^{iterative}(\alpha) \; .
\end{equation}
The function 
\begin{equation}
L_{2,2}^{sphere} = \frac{1+6 \ln 2}{4} G^\prime_2 - K^\prime
\end{equation}
occurs already in the expression (\ref{protoM}) for the magnetization 
$M_{sphere}(\alpha_s)$ of the sphere. The contribution
\begin{equation}
L_{2,2}^{iterative} = 64 {\cal L} ({\cal L}^{\prime})^2 + 
32 {\cal L}{\cal L}^{\prime \prime}
\end{equation}
\end{mathletters}
arises in obtaining the selfconsistent solution of the equation $M=M_{sphere}$
with an expansion and iteration.

\section{Discussion of the results} \label{discuss}

We will first show that our result (\ref{protoM}) for
$M_{sphere}(H_s)$ does not lead to a ferromagnetic solution in contradistinction
to the Weiss model. Then we discuss the behavior of the
different terms contributing to (\ref{MaginOphi1}) and to (\ref{MaginOphi2})
and we delineate the range of reliability of the simplest approximation.
Finally, we address problems arising when comparing with experiments. 

\subsection{Spontaneous magnetization?}
\label{ferromag}

Investigations based on density functional methods by Groh and Dietrich 
\cite{GD97a} and on Monte Carlo methods by Weis and Levesque
\cite{WL93a,WL93b} provided support for the existence of magnetized phases for 
absent {\em  external\/} field $H_e$, i.~e., ferromagnetism, in the system of 
dipolar hard spheres we consider in this work. 
Groh and Dietrich consider a ferrofluid probe of needle--like shape where 
$H = H_e$ and find a transition to a magnetized phase at 
$\phi \epsilon \approx 0.35$. But they consider this value as being 
overestimated and refer to \cite{WL93b}. Weis and Levesque study a case
without demagnetizing fields, i.~e., again $H = H_e$. They find a transition 
to a magnetized phase at $\epsilon = 6.25$ for $\phi \approx 0.35$. As 
discussed in detail below, these values are outside the range of reliability 
of our results.

The Weiss model does also show ferromagnetic behavior. It is recovered from 
(\ref{protoM}) by keeping only the leading--order term ${\cal L}(\alpha_s)$ 
describing a single moment in the field $H_s = H + M/3$. The resulting 
self--consistency equation
\begin{equation}
M = M^{Weiss}_{sphere} \left( H + \frac{M}{3} \right) = 
M_{sat} {\cal L} \left[ \frac{m}{kT} \left( H + \frac{M}{3} \right) \right]
\end{equation}
allows for zero field a solution with finite magnetization when 
$kT < m M_{sat}/9$. Using  (\ref{Msat}) combined with (\ref{phidef}) and
(\ref{epsidef}) this condition is equivalent to $\phi \epsilon > 3/8$, about
the same value as in \cite{GD97a}.
So according to the Weiss model the ferrofluid will show 
ferromagnetic behavior below a critical temperature that grows linearly with
the saturation magnetization $M_{sat}$ of the ferrofluid.
But even for a ferrofluid consisting of cobalt particles with a magnetic core
diameter of 10 nm and a magnetic volume fraction of $\phi_{mag} = 0.1$ the
critical temperature would be as low as 90 K.  

While the transition combination $\epsilon = 6.25$, $\phi \approx 0.35$ of 
\cite{WL93b} is outside the range of reliability of our results, the
threshold location $\phi \epsilon = 8/3$ 
of the Weiss model may be not. However, in agreement with \cite{WL93b} we do 
not find selfconsistent ferromagnetic solutions of the equation (\ref{protoM})
$M = M_{sphere}( H + M/3 )$ within this range.
We have numerically confirmed that for $H=0$ the equation
$M=M_{sphere}( M/3 )$ allows {\em always\/} only the trivial solution $M=0$.

\subsection{Contribution from different orders}
\label{contri}

Now we will take 
a closer look on the functions of $\alpha$ involved in (\ref{MaginOphi1}) and 
(\ref{MaginOphi2}). All these functions are odd as it has to be for reasons
of symmetry. For $\alpha \rightarrow \infty$ they vanish as $1/\alpha^2$
or faster. Because $1 - {\cal L}(\alpha) \sim 1/\alpha$ that means
that the predicted magnetization is always smaller than the saturation
magnetization for $\alpha \rightarrow \infty$. Nevertheless the magnetization
can assume unphysical values $> M_{sat}$ for intermediate $\alpha$ if 
$\epsilon$ or $\phi$ is big enough for the approximations to become invalid.

\subsubsection{Behavior in linear order of $\phi$}

We will first discuss the result (\ref{MaginOphi1}) for the magnetization that
was obtained up to linear order in the volume fraction $\phi$.
In figure \ref{first2inOphi} the functions $L_{1,1}$ and $L_{1,2}$ are
plotted. The values of the higher--order functions are smaller, but their shape
remains more or less the same as the logarithmic plot in 
figure \ref{allinOphi} shows.
Because $L_{1,n}$ and $L_{1,n+2}$ differ by about one order of magnitude 
one can conclude that by including higher and higher orders 
of $\epsilon$  the series (\ref{MaginOphi2}) for the magnetization converges, 
as long as 
$\epsilon$ is smaller than $\approx 3$. For this large value of $\epsilon$ 
strong agglomeration can already be expected.

For small $\alpha$, $L_{1,n}$ is proportional to $\alpha$ 
($\alpha^3$) for odd (even) $n$. The initial susceptibility can therefore
be written as
\begin{eqnarray}
&&\chi(H=0) = \nonumber \\
&& \qquad \chi_0(H=0) \left[ 1 + \phi \sum_{n=0} s_{1,2n+1} \epsilon^{2n+1} 
+ O(\phi^2) \right] \;\; .
\label{suscinOphi}
\end{eqnarray}
Here 
\begin{equation}
\chi_0(H=0) = \frac{m M_{sat}}{3 kT} \;\; ,
\end{equation}
is the initial susceptibility of the ideal paramagnetic gas,
and the nonvanishing $s_{1,n}$ we calculated are
\begin{eqnarray}
&& s_{1,1} = \frac{8}{3} ; \qquad s_{1,3} = \frac{8}{75}; \qquad
s_{1,5} = \frac{32}{3675} \nonumber \\
&& s_{1,7} = \frac{8}{19845}  ; \qquad s_{1,9} = \frac{148}{12006225} \;\; .
\end{eqnarray}

Figure \ref{suscinepsn} shows $\chi_0(H=0)$ (thick dashed line), and the 
susceptibility $\chi(H=0)$ (\ref{suscinOphi}) including progressive orders 
$\phi \epsilon$, $\phi \epsilon^3$, $\phi \epsilon^5$, $\phi \epsilon^7$, and 
$\phi \epsilon^9$ (thin dashed lines, from bottom to top) as a function of 
$\epsilon$ for $\phi = 0.15$. The sequence of these thin dashed lines shows 
that this series converges in the $\epsilon$--range of figure \ref{suscinepsn}.
The last thick full line in Figure \ref{suscinepsn} represents $\chi(H=0)$
including the contributions in  order $\phi^2 \epsilon^2$. It shows that the
latter are even for $\phi = 0.15$ not yet important.

\subsubsection{Behavior in second order of $\phi$}

Now we take a look at the functions $L_{2,2}^{sphere}$ (\ref{L22decomp}b) and 
$L_{2,2}^{iterative}$ (\ref{L22decomp}c) that add up to $L_{2,2}$ 
(\ref{L22decomp}a) which enters in
order $\phi^2 \epsilon^2$ into the magnetization (\ref{MaginOphi2}a).

Figure \ref{Ophi2} shows that the contributions $L_{2,2}^{sphere}$ and
$L_{2,2}^{iterative}$ almost cancel each other at small $\alpha$. This
is why the influence of the $\phi^2 \epsilon^2$--terms on the susceptibility in
figure \ref{suscinepsn} is so small. However, at higher $\alpha$ the 
$\phi^2 \epsilon^2 L_{2,2}$--term becomes important. Comparing the latter with 
the linear one, $\phi \epsilon L_{1,1}$, one finds that they contribute for
$\epsilon \phi \approx 0.5$ equally at larger $\alpha$.

Except for very small $\alpha$ $L_{2,2}$  is 
negative, because it includes higher--order particle position correlations
that result in a better modeling of the distance distribution due to the 
finite size of the particles. The mean distance is bigger in this approximation
and the induced dipolar fields at the particle positions are therefore smaller.

The influence of the $\phi^2 \epsilon^2 L_{2,2}$ contribution to the 
magnetization is shown in figure
\ref{magOphi2} for $\epsilon = 2$ and $\phi = 0.05$. For these
parameters this term is already
large enough to cancel almost exactly the sum of all contributions 
$L_{1,n} \phi \epsilon^n$, with $n \geq 2$ from the linear order in $\phi$
at moderate $\alpha$. Figure \ref{suszOphi2} shows the susceptibility 
$\chi(H) = \frac{\partial M(H)}{\partial H}$ for the
same parameters. At higher $\alpha$, the cancellation of the higher 
$L_{1,n}$--terms against the $L_{2,2}$--contribution can again be seen. At 
smaller $\alpha$, however, the behavior is different. There the contribution 
of the  $L_{1,n}$--terms is much larger, whereas the $L_{2,2}$--contributions 
vanish.

\subsection{Reliability of the $O(\phi \epsilon)$--approximation}

We can determine the range of reliability of the simplest approximation
\begin{eqnarray}
\frac{M}{M_{sat}} &=& L_{0,0}(\alpha) + L_{1,1} (\alpha) \phi \epsilon 
\nonumber \\
&=& {\cal L}(\alpha) + 8 {\cal L}(\alpha) {\cal L}'(\alpha) \phi \epsilon
\label{simpleM}
\end{eqnarray}
to the magnetization that includes effects of dipolar interactions since we know
the higher--order corrections in $\phi$ as well as in $\epsilon$. To that end 
we investigated the ratios
\begin{equation}
\left| \frac{ O(\phi \epsilon^n)\text{--terms ($n > 1$)}}
{L_{0,0}(\alpha) + L_{1,1} (\alpha) \phi \epsilon}\right| \;\; ,
\label{epsnerror}
\end{equation}
and 
\begin{equation}
\left| \frac{ O(\phi^2)\text{--terms}}
{L_{0,0}(\alpha) + L_{1,1} (\alpha) \phi \epsilon} \right|\;\; .
\label{phi2error}
\end{equation}
The first ratio assumes its maximum at $\alpha = 0$, that means the initial
susceptibility is most sensitive to higher order corrections in $\epsilon$. 
The second ratio (\ref{phi2error}) assumes its
maximum around $\alpha = 2$, that is near the maximum of the absolute value of
the numerator (as seen in figure \ref{Ophi2}). 

In the $\epsilon$--$\phi$ plane of figure \ref{errors}(a) we show isolines
of the maximal -- with respect to $\alpha$ -- ratio (\ref{epsnerror}) and
figure \ref{errors}(b) shows the analogous isolines for the ratio 
(\ref{phi2error}). 
The comparison shows that the smallness of
$\epsilon$ is more important to keep the ratio (\ref{epsnerror}) small,
whereas in (\ref{phi2error}) the value of $\phi$ is also important. As rules of
thumb one can say that the approximation (\ref{simpleM}) is valid within 
about 1 -- 2 percent if $\epsilon < 1$ and $\epsilon \phi < 0.04$. If the first
constraint is not fulfilled, higher orders in $\epsilon$ have to be taken
into account. Higher orders in $\phi$ are needed if the second constraint is 
not fulfilled.

\subsection{Comparison with experiments?}\label{compari}

There are several papers \cite{P95,ML90,SPMS90,PML} that aim at investigating 
the influence of dipolar interactions on the magnetization by comparing
theoretical models developed so far with experimental magnetizations of
ferrofluids. The mean spherical model \cite{W71} was reported
to show good agreement with experiments. Pshenichnikov \cite{P95}
found also good agreement with the high temperature 
approximation \cite{BI92}, i.~e., the approximation (\ref{simpleM}). But 
this ansatz failed in the magneto--granulometric analysis done in  \cite{PML}.

We do not present a comparison of our results with the experiments on the 
magnetization in the literature because of several problems:
In our theory it is necessary to distinguish between the 
particle diameter $D$ and the magnetic core diameter $D_{mag}$ that is found
in magneto--granulometric measurements. This problem does not arise in 
the mean spherical model or the high temperature approximation, where 
$\epsilon$ and $\phi$ enter only via the factor 
$\phi \epsilon = \frac{N m^2}{24 V \mu_0 k T} = \frac{M_{sat} m}{24 kT}$ 
that is independent of $D$. Also, corrections such as the temperature 
dependence of the saturation magnetization or the fluid density should
be taken into account \cite{SPMS90}.  

But the major problem in comparing directly with experiments is that our theory
does not take into account the polydispersity of ferrofluids. The effect of
polydispersity is significant already in the absence of any dipolar interaction.
This can be inferred from the dashed and the dotted curves in Fig.~\ref{polymag}
representing the reduced magnetization of {\em noninteracting\/} magnetic
particles having a polydisperse and a monodisperse distribution of particle
diameters, respectively. Here the common particle diameters of the latter is
$\overline{D^3}^{1/3}$, where $\overline{D^3}$ is the third moment of the
particle size distribution
\begin{equation}
P(D) := \frac{1}{\sqrt{ 2 \pi} \sigma D_0 e^{\sigma^2/2}} 
e^{-\frac{\ln^2 D/D_0}{ 2 \sigma^2}} \;\; .
\label{lognormal}
\end{equation}
of the former.
The mean magnetic moment $\overline{m}$ and the saturation 
magnetization of the two systems are the same. For comparison with the effect of
dipolar interaction in {\em monodisperse\/} systems the full curve in 
Fig.~\ref{polymag} shows our result for $M$ (\ref{MaginOphi2}a) including all
terms $\sim \phi \epsilon^n$ and the term $\phi^2 \epsilon^2$. Hence the effects
of polydispersiveness alone, i.~e., without interaction are comparable in size
with the effect of dipolar interactions in monodisperse systems. Thus clearly an
extension of the here presented Born--Mayer expansion method to the case of
polydisperse interacting particles is desirable.

\section{Conclusion}
\label{conclude}

We calculated the free energy and in particular the magnetization $M$ of a 
ferrofluid as a function of the macroscopic magnetic field $H$.  To do so, 
we used the technique of the Born--Mayer expansion together with an expansion 
in terms of the dipolar coupling energy. The magnetic particles were assumed to 
be hard spheres with a common hard core diameter $D$ and magnetic moment $m$
that interact via long range dipolar interactions. This feature may result in a
geometry dependence of thermodynamic properties. We treated this problem by
dividing the dipolar field at some position ${\bf x}_i$ that is produced by the
magnetic moments of the particles into a near--field and into a far--field part
depending on  whether the particle distance from ${\bf x}_i$ is larger than some
radius $R_s$ or not. In this way {\em every\/} magnetic particle is imagined to
be located in the center of a sphere of radius $R_s$. The far--field dipolar
contribution from particles beyond $R_s$ is then replaced by a magnetic
continuum with magnetization $M$ and magnetic field $H$. 
Here $R_s$ is chosen to be such that $M$ and $H$ are homogeneous on the scale of
$R_s$.
The magnetic continuum
outside the sphere produces in the center of the sphere the magnetic field $H_s
= H + M/3$. This field acts as an "external" field on the particle in the center
of the sphere. The near--field interaction of the latter with the other
particles within the sphere being at a distance smaller than $R_s$ is treated
explicitly. Thus in our statistical mechanical calculations there appear dipolar
interactions only with interparticle distances less than $R_s$. 
However, since the cutoff dependence of the relevant expressions occurring in
these calculations is negligible already beyond a radius of the order of $10 D
\approx 100$ nm we used $R_s = \infty$ in these expressions.

The expansion of the partition function for these interacting particles in terms
of the volume ratio and the dipolar coupling strength $\epsilon$ yields an
expression for the magnetization
\begin{equation}
M_{sphere} = M_{sphere} ( H + M/3) \;\; .
\end{equation}  
as a function of the "external" part of the field inside the sphere. 
The magnetization $ M_{sphere}$ is then identified with the magnetization 
$M(H)$ inside the continuum so that a selfconsistent relation results.
The aforementioned geometry dependence of $M$ in the general case is
incorporated via $H$.
  
We presented two different expansions in $\epsilon$ and $\phi$, one containing
only linear terms in $\phi$, the other also second order $\phi$ terms, but only 
up to $O(\epsilon^2)$. We  discussed the range of applicability in the 
$\phi$--$\epsilon$ plane of their results for $M(H)$ and compared them to the
most simple approximation to the magnetization that contains the dipolar
effects only in linear order in $\epsilon$ and $\phi$.
The selfconsistent relation for $M(H)$ that contains only up to second order 
terms in both parameters does not admit a ferromagnetic solution with
spontaneous magnetization. Finally we showed that an extension to polydisperse
interacting particles is desirable.

\acknowledgments

This work was supported by the Deutsche Forschungsgemeinschaft (SFB 277).

\appendix

\section{The functions $G_\char110$ and $K$}
\label{appa}

The functions $G^*_n(x)$ in (\ref{lastintegral}) are related to 
$G_n(x)$ via (\ref{Gdef}):
\begin{equation}
G_n(x) = \frac{1}{16 \pi^3 (n-1) n!} 
\left( \frac{x}{\sinh x} \right)^2 G^*_n(x) \;\; . 
\end{equation} 

The functions $G_n(x)$ introduced in (\ref{Gdef}) have the form
\begin{eqnarray}
G_n(x) &=& G_n^{(0)}\left( \frac{1}{x} \right) + 
G_n^{(1)} \left( \frac{1}{x} \right) \coth x \\
&& + G_n^{(2)} \left( \frac{1}{x} \right) \coth^2 x
\nonumber \;\; ,
\end{eqnarray}
where the functions $G_n^{(i)}(y)$ are polynomials. The first four triple are 
given by
\begin{mathletters}
\begin{eqnarray}
G_2^{(0)}(y) &=& \frac{8}{5} + \frac{8}{5} y^2 +\frac{12}{5} y^4 \\
G_2^{(1)}(y) &=& - \frac{8}{5} y - \frac{24}{5} y^3  \\
G_2^{(2)}(y) &=& \frac{12}{5} y^2 
\end{eqnarray}
\end{mathletters}
\begin{mathletters}
\begin{eqnarray}
G_3^{(0)}(y) &=& -\frac{4}{35} y^2 - \frac{48}{35} y^4 -\frac{12}{7} y^6 \\
G_3^{(1)}(y) &=& -\frac{8}{35} y + \frac{8}{5} y^3 +\frac{24}{7} y^5 \\
G_3^{(2)}(y) &=& \frac{16}{105} - \frac{8}{35} y^2 - \frac{12}{7} y^4 
\end{eqnarray}
\end{mathletters}
\begin{mathletters}
\begin{eqnarray}
G_4^{(0)}(y) &=& 
\frac{8}{105} + \frac{8}{35} y^2 + \frac{92}{35} y^4 + \frac{72}{7} y^6 + 12 y^8\\
G_4^{(1)}(y) &=& 
-\frac{16}{105} y- \frac{8}{5} y^3 - \frac{88}{7} y^5 -24 y^7\\
G_4^{(2)}(y) &=&
\frac{32}{105} y^2 + \frac{16}{7} y^4 +12 y^6 
\end{eqnarray}
\end{mathletters}
\begin{mathletters}
\begin{eqnarray}
G_5^{(0)}(y) &=& 
\frac{12}{385} y^2 - \frac{208}{385} y^4 - \frac{852}{77} y^6 - 
\frac{480}{11} y^8 - \frac{540}{11} y^{10} \\
G_5^{(1)}(y) &=& 
-\frac{8}{231} y + \frac{16}{385} y^3 + \frac{472}{77} y^5 + 
\frac{600}{11} y^7 + \frac{1080}{11} y^9 \\
G_5^{(2)}(y) &=& \frac{16}{1155} + \frac{16}{1155} y^2 - 
\frac{40}{77} y^4 - \frac{120}{11} y^6 - \frac{540}{11} y^8 \;\; .
\end{eqnarray}
\end{mathletters}
All functions $G_n^{(i)}(x)$ have a well defined limit for $x \rightarrow 0$
although this is not obvious for the above explicit expressions. Their values 
at $x = 0$ are closely related to the coefficients in the $\epsilon$--expansion 
of the second virial coefficient for the system of dipolar hard spheres in the 
absence of a magnetic field. The calculation of this coefficient dates back to
\cite{Ke12} and can also be found in \cite{BH76}.

The function $K$ (\ref{Kdef}) that appears in
the $O(\phi^2)$--terms of the free energy is given by
\begin{eqnarray}
K(x) &=& -\frac{6}{x} 
\coth^3 x + \left( \frac{18}{x^2} + 12 \right) \coth^2 x \nonumber \\
&-& 
\left( \frac{18}{x^3} + \frac{24}{x} \right) \coth x + 
\frac{6}{x^4} + \frac{12}{x^2} \;\; . \label{Kexpr}
\end{eqnarray}

\section{Graphs in second order of $\phi$}\label{appb}

Here we determine the contribution to the canonical partition function from 
the graphs A--H shown in Fig.~\ref{clusterexp2}. There often appear hard core
interaction terms that are just expressions of the requirement that two 
particles have to or must not overlap. We define two abbreviations:
\begin{equation}
e^{-v^{HC}_{ij}} - 1 = - O_{ij} \;\; , 
\end{equation}
\begin{equation}
e^{-v^{HC}_{ij}} = \overline{O}_{ij} \;\; . 
\end{equation}

\subsection{Graph A}

The graph A represents $f^{(0)}_{ij} f^{(0)}_{ik}$. 
There are $N^3$ ways to choose the constituting particles. But because $j$ and 
$k$ are equivalent only $N^3/2$ distinctive graphs remain. Integration over all
variables except the positions of particle $j$ and $k$ relative to $i$ yields
\begin{eqnarray}
&& \frac{N^3}{2}\int f^{(0)}_{12} f^{(0)}_{13} \prod_l e^{-v_l} 
\; d\stackrel{\rightarrow}{\bf x} d\stackrel{\rightarrow}{\Omega}
\nonumber \\
&=& \frac{N^3}{2} \left( 4 \pi \frac{\sinh \alpha_s}{\alpha_s} \right)^N 
V^{N-2} \int  O_{12} O_{13} d {\bf r}_{12} d {\bf r}_{13} \;\; . \nonumber
\end{eqnarray}  
The remaining integral factorizes and we can make use of the results
for $A_0$ (\ref{A0result}). The contribution of graph A to the 
partition function is
\begin{equation}
Z_A = 32 N Z_0 
\phi^2
\;\; 
\end{equation}
where $Z_0$ is given by (\ref{Z0def}).

\subsection{Graph B}

The graph B represents $f^{(0)}_{ij} f^{(2)}_{ik}$. All three particles appear 
in different ways, thus there are $N^3$ different graphs. After integration over
the degrees of freedom of the noninvolved particles and switching to 
relative coordinates with respect to particle $i$ the integral factorizes
again and one can make use of the results for $A_0$ (\ref{A0result}) and 
$A_2$ (\ref{Aninf}). We get
\begin{equation}
Z_B = -16 N Z_0 
\phi^2 \epsilon^2 G_2(\alpha_s) \;\; .
\end{equation}

\subsection{Graph C}

The graph C represents $f^{(0)}_{ij} f^{(0)}_{kl}$. Here we have also to
include the next higher order term when we calculate the number of combinations
to get the $O(N)$--terms in the final result:
There are $(N^4 - 6 N^3)/8$ different terms. The integral for graph C can be 
factorized so that
\begin{equation}
Z_C = 8( N^2 - 6 N) Z_0 \phi^2 \;\; .
\end{equation}

\subsection{Graph D}

The graph D represents $f^{(0)}_{ij} f^{(2)}_{kl}$. The calculation is similar
to the calculation of graph C. Again, we need the next higher order term in 
$N$. There are $(N^4 - 6 N^3)/4$ combinations, twice as much as for graph C
because the pairs $(i,j)$ and $(k,l)$ are not identical. One gets
\begin{equation}
Z_D = -4  ( N^2 - 6 N) Z_0 \phi^2 \epsilon^2 
G_2(\alpha_s) \;\; .
\end{equation}

\subsection{Graph E}

The integral containing the term $f^{(0)}_{ij} f^{(0)}_{jk} f^{(0)}_{ki}$ is 
the first really new integral. It involves only hard core interactions and does
not contribute to the final expression for the magnetization. But for 
completeness we will calculate it also. The trivial integrations yield
\begin{equation}
Z_E = -\frac{N^3}{6} 
\left( 4 \pi \frac{\sinh \alpha_s}{\alpha_s} \right)^N V^{N-2}
\int O_{12} O_{13} O_{23} d {\bf r}_{12} d {\bf r}_{13} \;\; .
\end{equation}
We keep the distance ${\bf r}_{12}$ fixed. The center of particle 3 has then 
to be inside two spheres of radius $D$ around particle 1 and 2. Integrating 
over the position of particle 3 yields the overlap volume $V_o$ of the 
two spheres
\begin{equation}
V_o = \frac{4}{3} \pi D^3 \left[ 1 - \frac{3}{4} \frac{r_{12}}{D} + 
\frac{1}{16} \left( \frac{r_{12}}{D} \right)^3 \right] \;\; .
\end{equation}  
Therefore
\begin{eqnarray}
Z_E &=& -\frac{N^3}{6} \frac{Z_0}{V^2}
\int  O_{12} V_o(r_{12}) d {\bf r}_{12} \\
&=& -\frac{N^3}{6}  \frac{Z_0}{V^2}
4 \pi \int_0^D  V_o(r_{12}) r^2_{12} d r_{12}\;\; .\nonumber 
\end{eqnarray}
Performing the last integration results in
\begin{eqnarray}
Z_E &=& -5 N Z_0 
\phi^2 \;\; .
\end{eqnarray}

\subsection{Graph F}

The graph F represents $f^{(0)}_{ij} f^{(0)}_{ik} f^{(1)}_{jk}$.
As already stated this integral vanishes which can be seen as follows: 
Consider an arbitrary configuration belonging to some value of the integrand
\begin{equation}
e^{-v_i - v_j - v_k} f^{(0)}_{ij} f^{(0)}_{ik} f^{(1)}_{jk} \;\; .
\end{equation}
While leaving the direction of the magnetic moments fixed the whole
configuration can be freely rotated around particle $j$ changing only the
$f^{(1)}_{jk}$ term. Integration over the resulting configurations involves
again an averaging over a dipolar field on a spherical surface. 
  
\subsection{Graph G}

The calculation of the $N^3/2$ integrals belonging to 
$f^{(0)}_{ij} f^{(0)}_{ik} f^{(2)}_{jk}$ is similar to the calculation for graph
E. First we integrate over all degrees of freedom except the distance between
particle $j=1$ and $k=2$ and the position of particle $i=3$:
\begin{eqnarray}
Z_G &=& \frac{N^3}{2} 
\left( 4 \pi \frac{\sinh \alpha_s}{\alpha_s} \right)^N V^{N-2}
\pi G_2(\alpha_s) \epsilon^2 D^6 \nonumber \\
&& \times \int  O_{13} O_{23} \overline{O}_{12} r_{12}^{-4} d r_{12} d{\bf r}_3 \;\; .
\end{eqnarray}
Integrating over ${\bf r}_3$ results again in an overlap volume term:
\begin{equation}
Z_G = \frac{N^3}{2} \frac{Z_0}{V^2}
\pi G_2(\alpha_s) \epsilon^2 D^6
\int \overline{O}_{12} V_o(r_{12}) r_{12}^{-4} d r_{12} \;\; . 
\end{equation}
Here the lower integration boundary is
$r_{12} = D$ because of the remaining hard core factor. The upper
integration boundary is $r_{12} = 2D$ because the possibility 
that particle $3$ overlaps with particle $1$ and $2$ is still required. The
final result is
\begin{equation}
Z_G = \frac{1 + 6 \ln 2}{4} N Z_0
\phi^2 \epsilon^2 G_2(\alpha_s) \;\; .
\end{equation}
 
\subsection{Graph H}

The last graph H is the most complicated one. It represents the term
$f^{(0)}_{ij} f^{(1)}_{ik} f^{(1)}_{jk}$ that appears $N^3/2$ times. The 
problem here
is to fulfill the requirement that particles $j$ and $k$ have to overlap in 
terms of properly chosen integration limits. We start with performing 
the trivial integrations:
\begin{eqnarray}
Z_{H} &=&  -\frac{1}{2} N^3 z_0^{N-3} V  \nonumber \\ 
&& \times \int e^{-v_1 -v_2 -v_3} v_{12}^{DD} v_{13}^{DD} 
O_{23} \overline{O}_{12} \overline{O}_{13} 
 \label{ZH1} \\
&& \times d{\bf r}_{12} d{\bf r}_{13} d \Omega_1 d \Omega_2 d \Omega_3\;\; . 
\nonumber 
\end{eqnarray}
Whether the integrand vanishes due to the hard core factors depends only on 
the distances $r_{12}$, $r_{13}$, and the angle $\vartheta_{23}$ between  
${\bf r}_{12}$ and ${\bf r}_{13}$. Consider a special orientation where
\begin{eqnarray}
{\bf \hat{r}}^0_{12} &=& (1,0,0) \\
{\bf \hat{r}}^0_{13} &=& (\cos \vartheta_2, 0, \sin \vartheta_{23}) \;\; ,
\end{eqnarray}
with $0 \leq \vartheta_{23} \leq \pi$.
A general configuration of the particles' locations can be written as
\begin{equation}
{\bf \hat{r}}_{1(2,3)} = {\cal R}_z(\varphi) 
{\cal R}_y(\vartheta) {\cal R}_z(\psi) 
{\bf \hat{r}}_{1(2,3)}^0 \;\; ,
\end{equation}
where ${\cal R}_x$, ${\cal R}_y$, and ${\cal R}_z$ are Eulerian rotation 
matrices for the angles
$\psi$, $\vartheta$, and $\varphi$. Using this form the integration over the
factors that depend on these angles:
\begin{equation}
\int_0^{2 \pi} \int_{-\pi/2}^{\pi/2} \int_0^{2 \pi} 
v_{12}^{DD} v_{13}^{DD}
\cos \vartheta d\varphi  d \vartheta d\psi \;\; ,
\end{equation}
can easily be performed with {\em mathematica\/}. We call the result
$I(r_{12},r_{13},\vartheta_{23},{\bf m}_i)$.

Next, we integrate over the orientations of the ${\bf m_i}$:
\begin{equation}
\int e^{-v_1-v_2-v_3} I(r_{12},r_{13},\vartheta_{23},{\bf m}_i) 
d \Omega_1 d \Omega_2 d \Omega_3 \;\; .
\end{equation}
The result depends only on $r_{12}$, $r_{13}$, and $\vartheta_{23}$. 
Using it in (\ref{ZH1}) yields
\begin{eqnarray}
Z_{H} &=&  -\frac{4}{3} N^3 \frac{Z_0}{V^2} K(\alpha_s) \nonumber \\ 
&& \times \int O_{23} \overline{O}_{12} \overline{O}_{13}
\frac{m^4 \pi^2 (2 \cos^2 \vartheta_{23} - \sin^2 \vartheta_{23})}
{10 (4 \pi \mu_0 k T)^2  r_{12} r_{13}} \nonumber  \\
&& \times \sin \vartheta_{23} dr_{12} dr_{13} d \vartheta_{23} \;\; , 
\label{ZH2}
\end{eqnarray}
where
\begin{eqnarray}
K(\alpha_s) &=& \frac{3}{8} \left( \frac{\alpha_s}{\sinh \alpha_s} \right)^{3} 
\int_{-1}^1 \int_{-1}^1 \int_{-1}^1
e^{\alpha_s (u_1 + u_2 + u_3)} \nonumber \\
&& \times (u_1^2 + 3) u_2 u_3 d u_1 d u_2 d u_3 \;\; . \label{Kdef}
\end{eqnarray}
The explicit expression for $K(\alpha_s)$ is given in Appendix \ref{appa}
(eqn.~\ref{Kexpr}).

Now we discuss the hard core terms. $r_{12}$ and $r_{13}$
have to be greater than $D$ to avoid the overlap with particle $1$. Furthermore
$|r_{12} - r_{13}| < D$ has to be fulfilled for particle 2 and 3 to overlap.
As a last requirement, $\vartheta_{23}$ has to be smaller than some angle
$\vartheta^{\text{max}}_{23}$ that depends on $r_{12}$ and $r_{13}$.
Trigonometry shows that
\begin{equation}
\cos \vartheta^{\text{max}}_{23} = 
\frac{r_{12}^2 + r_{13}^2 - D^2}{2 r_{12} r_{13}} \;\; .
\end{equation}
In this configuration the distance between particle 2 and 3 is exactly $D$.

We perform the integration over $\vartheta_{23}$ from $0$ to 
$\vartheta^{\text{max}}_{23}$ in (\ref{ZH2}), choose the correct limits
for $r_{12}$ and $r_{13}$, and drop all hard core terms: 
\begin{eqnarray}
Z_{H} &=&  -\frac{4}{3} N^3 \frac{Z_0}{V^2} K(\alpha_s) \nonumber \\ 
&& \times \int_D^\infty \int_{\min(D,r_{13}-D)}^{r_{13}+D} 
\frac{m^4 \pi^2}{10 (4 \pi \mu_0 k T)^2  r_{12} r_{13}} \\
&&  \times \left[  \frac{r_{12}^2 + r_{13}^2 - D^2}{2 r_{12} r_{13}} - 
\left( \frac{r_{12}^2 + r_{13}^2 - D^2}{2 r_{12} r_{13}} \right)^3
\right] dr_{12} dr_{13}  \nonumber \;\; , 
\end{eqnarray}
The result of the last integration is:
\begin{eqnarray}
Z_{H} &=&  -\frac{4}{3} N^3 \frac{Z_0}{V^2} K(\alpha_s)
\frac{m^4 \pi^2} {48 (4 \pi \mu_0 k T)^2} \nonumber \\
&=& -N Z_0
\phi^2 \epsilon^2 K(\alpha_s) \;\; . 
\end{eqnarray}

\newpage
\narrowtext

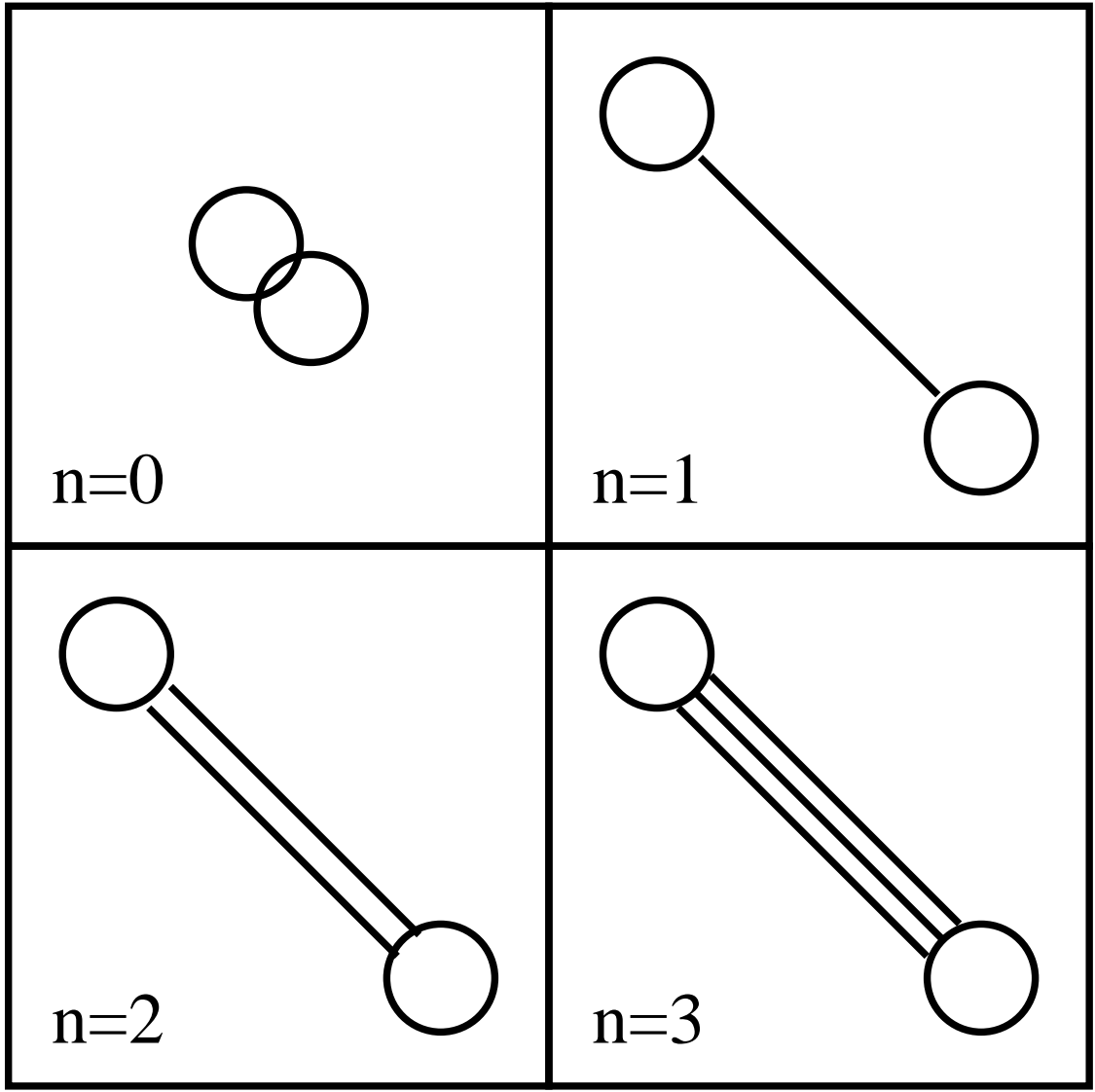
\begin{figure}\caption[]
{The first four graphs needed for the expansion of $Z$ in 
section~\ref{exp1}. They correspond to the terms $f^{(n)}_{ij}$ 
($n=0,1,2,3$), and are of the order
$\phi \epsilon^n$.}
\label{clusterexp1}
\end{figure}
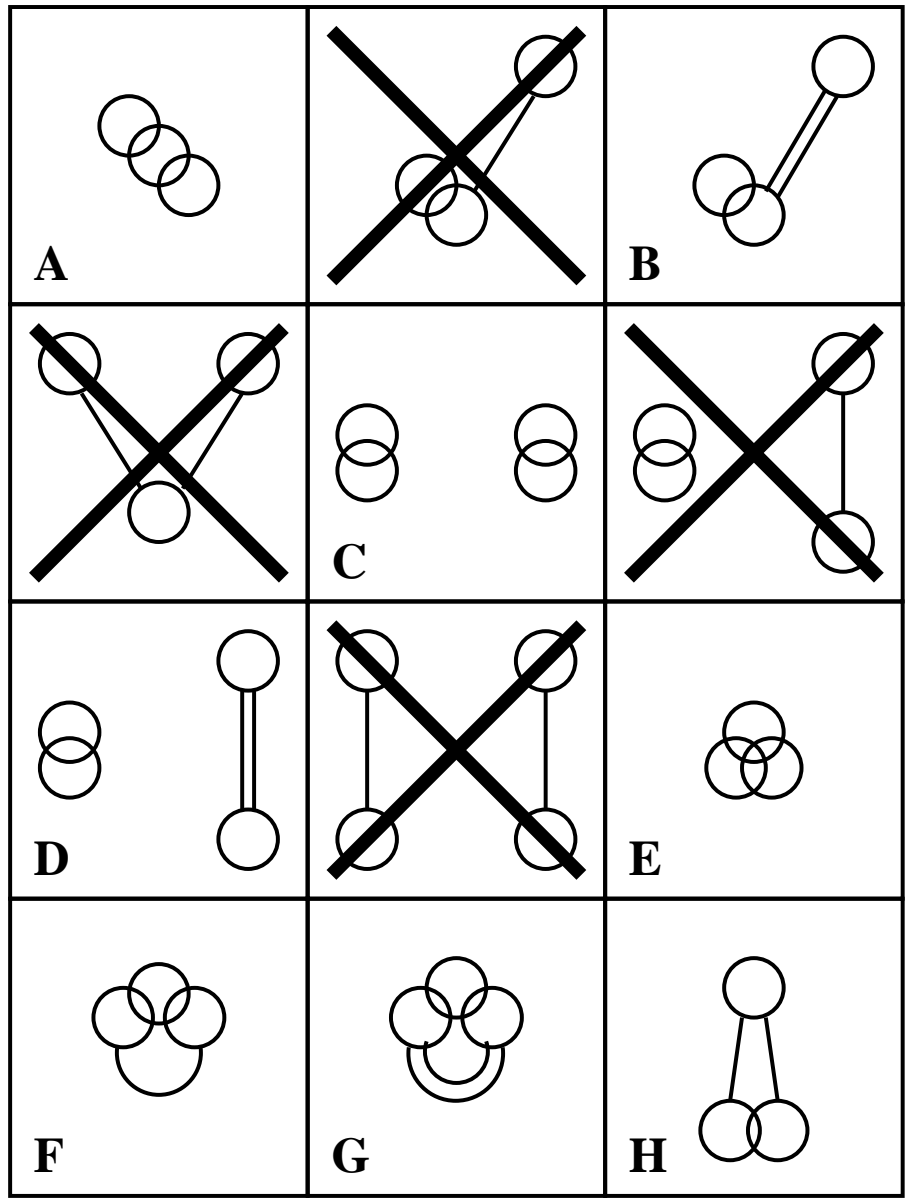
\begin{figure}\caption[]
{The twelve additional graphs needed for an $O(\phi^2, \epsilon^2)$ expansion 
of $Z$. The integrals for the crossed out graphs vanish. Graph F vanishes also;
see Appendix~\ref{appb}6.}
\label{clusterexp2}
\end{figure}
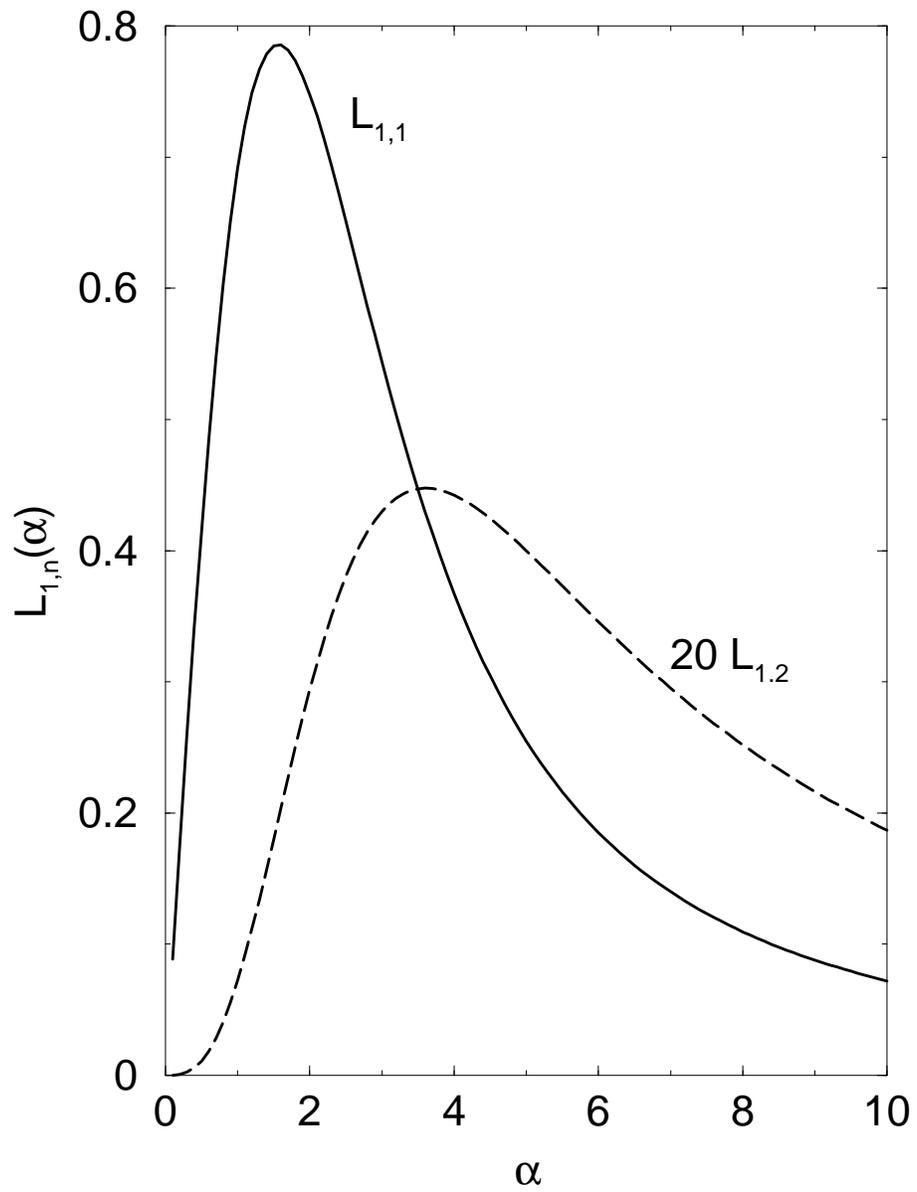
\begin{figure}\caption[]
{The functions $L_{1,1}$ and $L_{1,2}$ versus $\alpha$. Note the different 
scaling.}
\label{first2inOphi}
\end{figure}
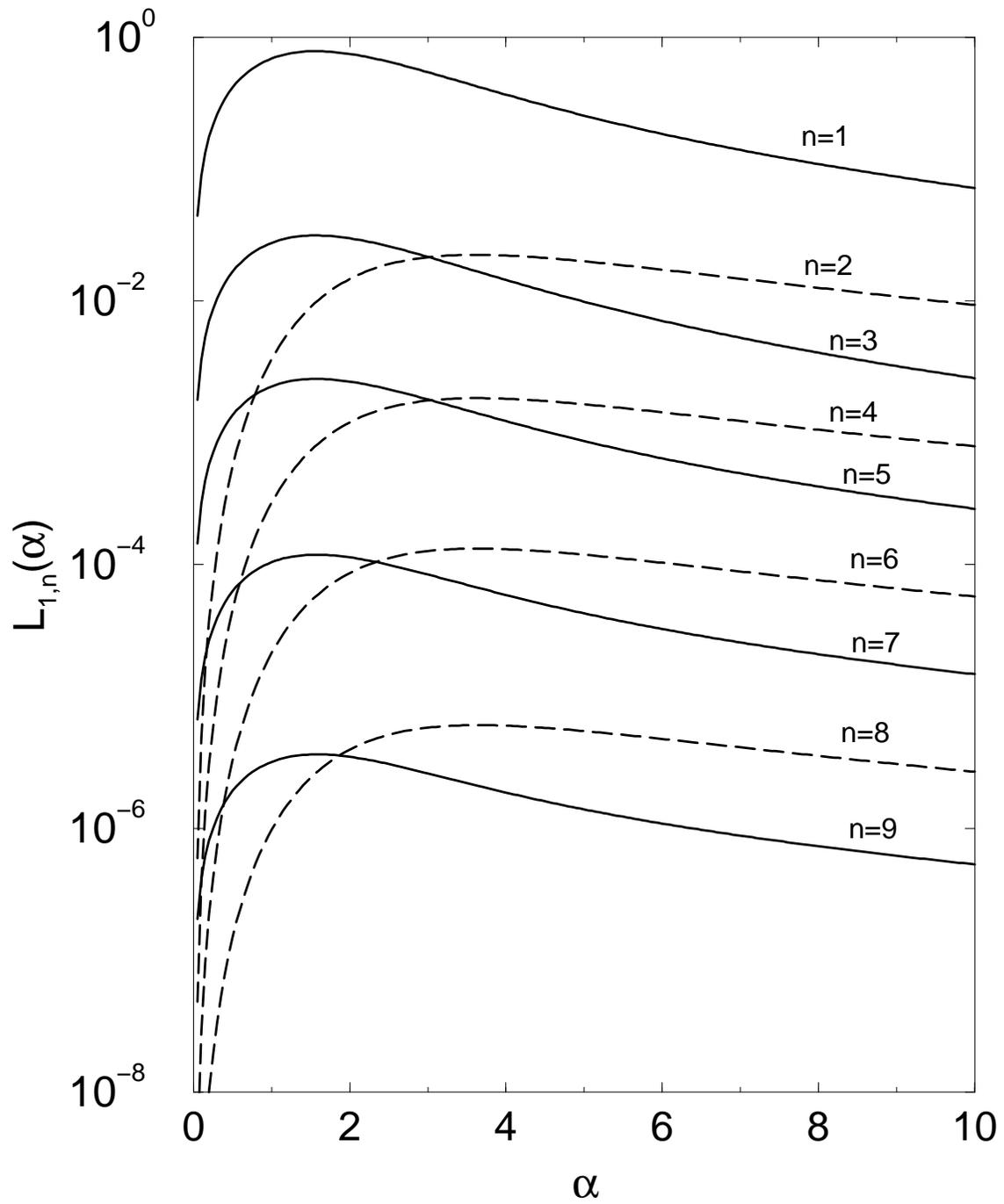
\begin{figure}\caption[]
{The functions $L_{1,n}$ versus $\alpha$.}
\label{allinOphi}
\end{figure}
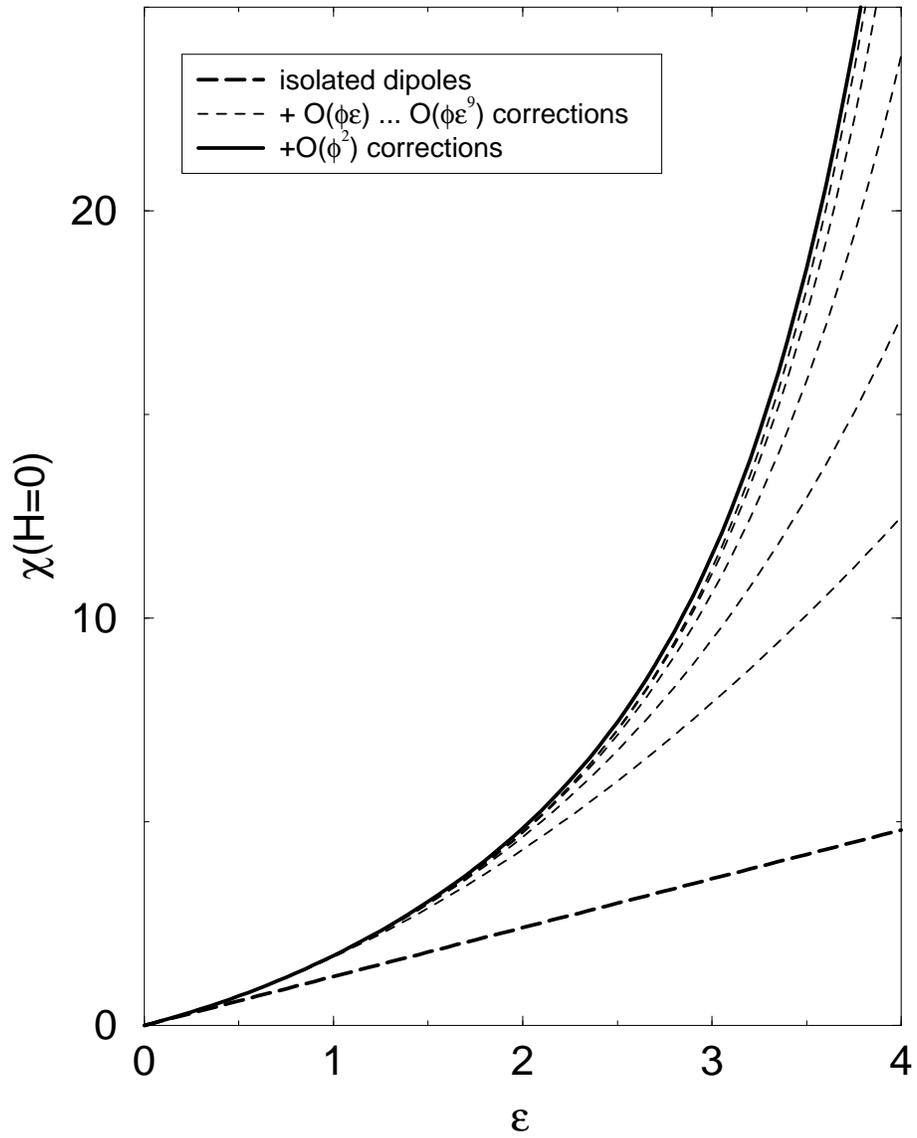
\begin{figure}\caption[]
{Initial magnetic susceptibility for $\phi = 0.15$ as a function of $\epsilon$.}
\label{suscinepsn}
\end{figure}
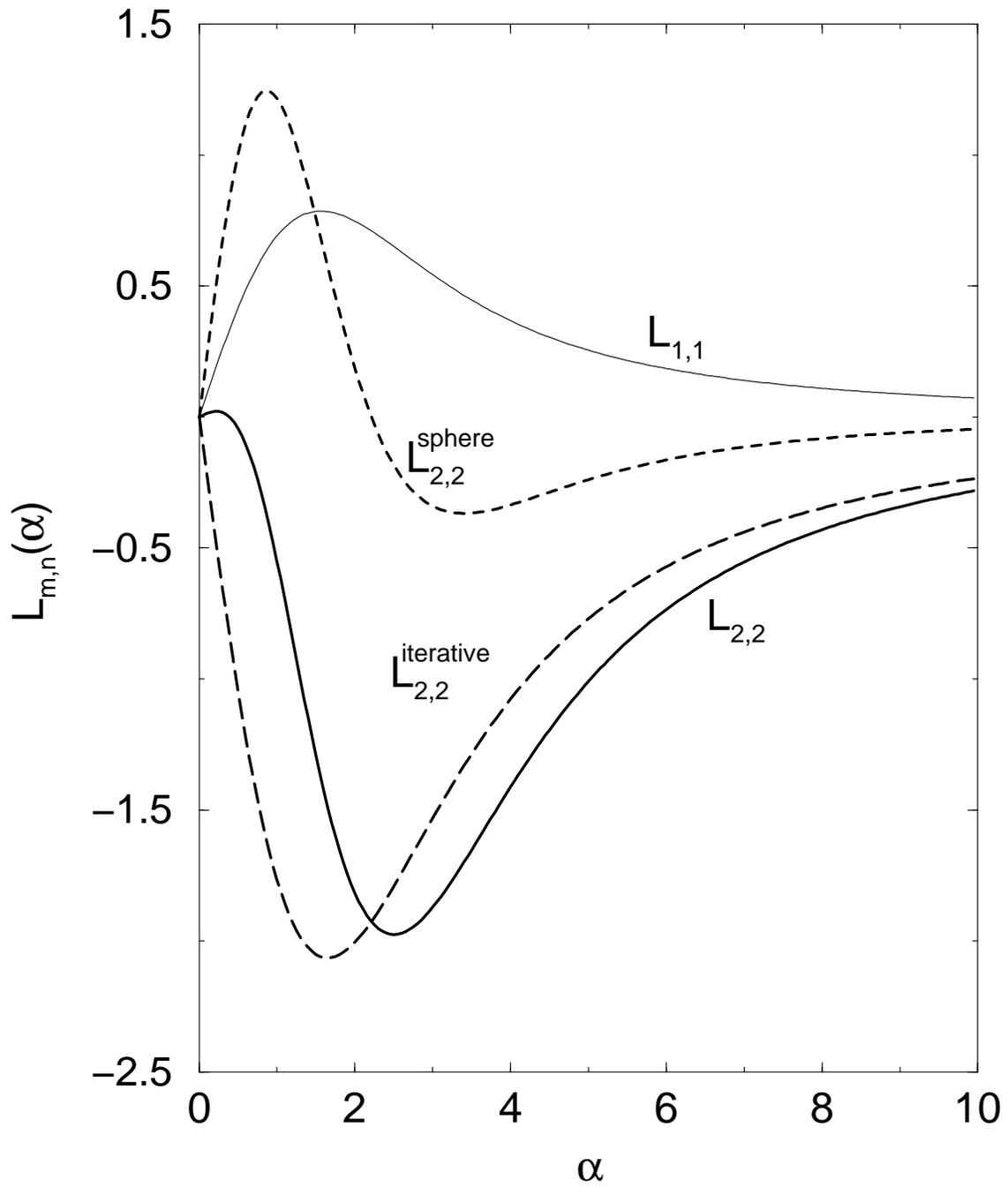
\begin{figure}\caption[]
{The weight of the $O(\phi^2)$--terms that appear in (\ref{MaginOphi2}) are
shown versus $\alpha$. The terms $L_{2,2}^{sphere}$ (\ref{L22decomp}b) and
$L_{2,2}^{iterative}$ (\ref{L22decomp}c) that add up to $L_{2,2}$
(\ref{L22decomp}a) are discussed in the text. For comparison, the 
$O(\phi \epsilon)$--term $L_{1,1}$ is plotted as well.}
\label{Ophi2}
\end{figure}
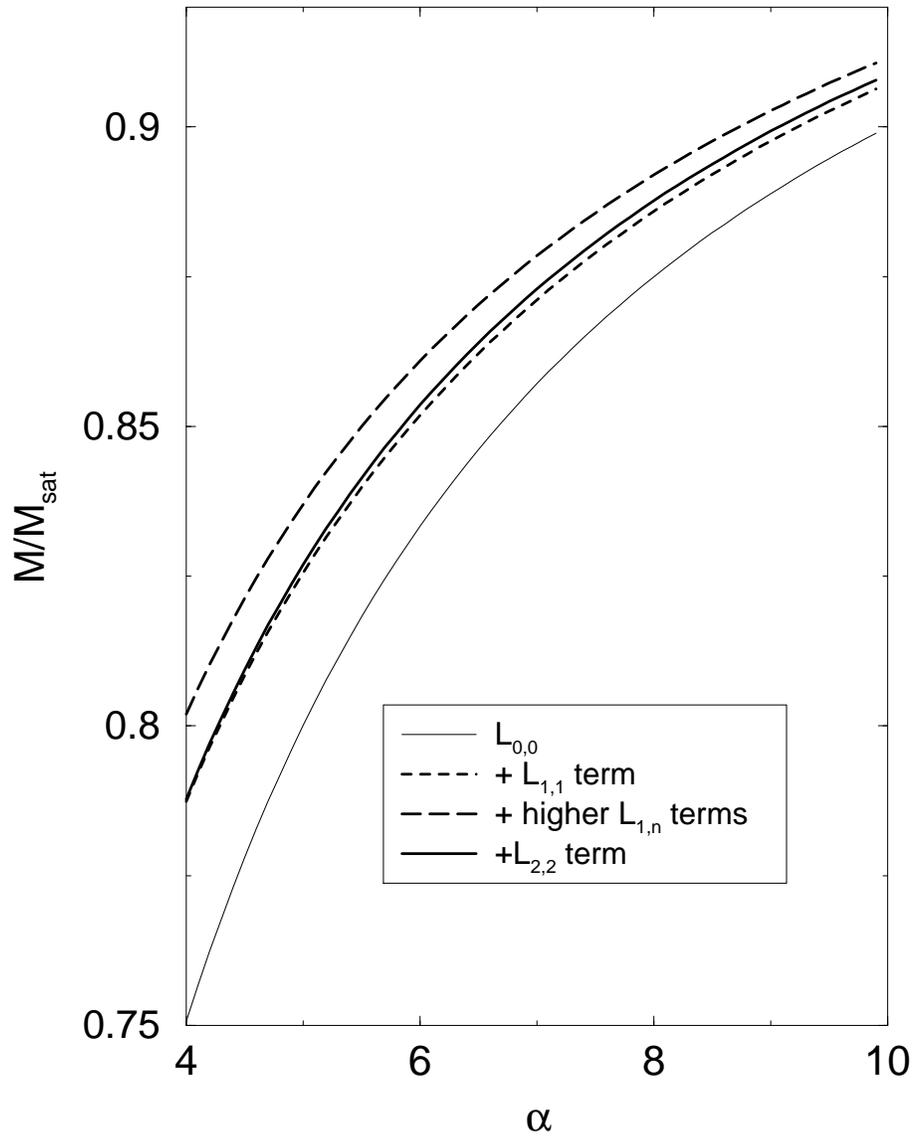
\begin{figure}\caption[]
{The reduced magnetization for $\epsilon=2$ and
$\phi = 0.05$ for moderate $\alpha$. Taking into account 
$\phi \epsilon^n$--terms results in a higher magnetization than 
given by the Langevin function. However, all contributions from the terms 
$\phi \epsilon^n$ with $n \geq 2$ are almost exactly canceled by the
contribution from the second order term $\phi^2 \epsilon^2$ for the parameters 
$\epsilon$, $\phi$ considered here.}
\label{magOphi2}
\end{figure}
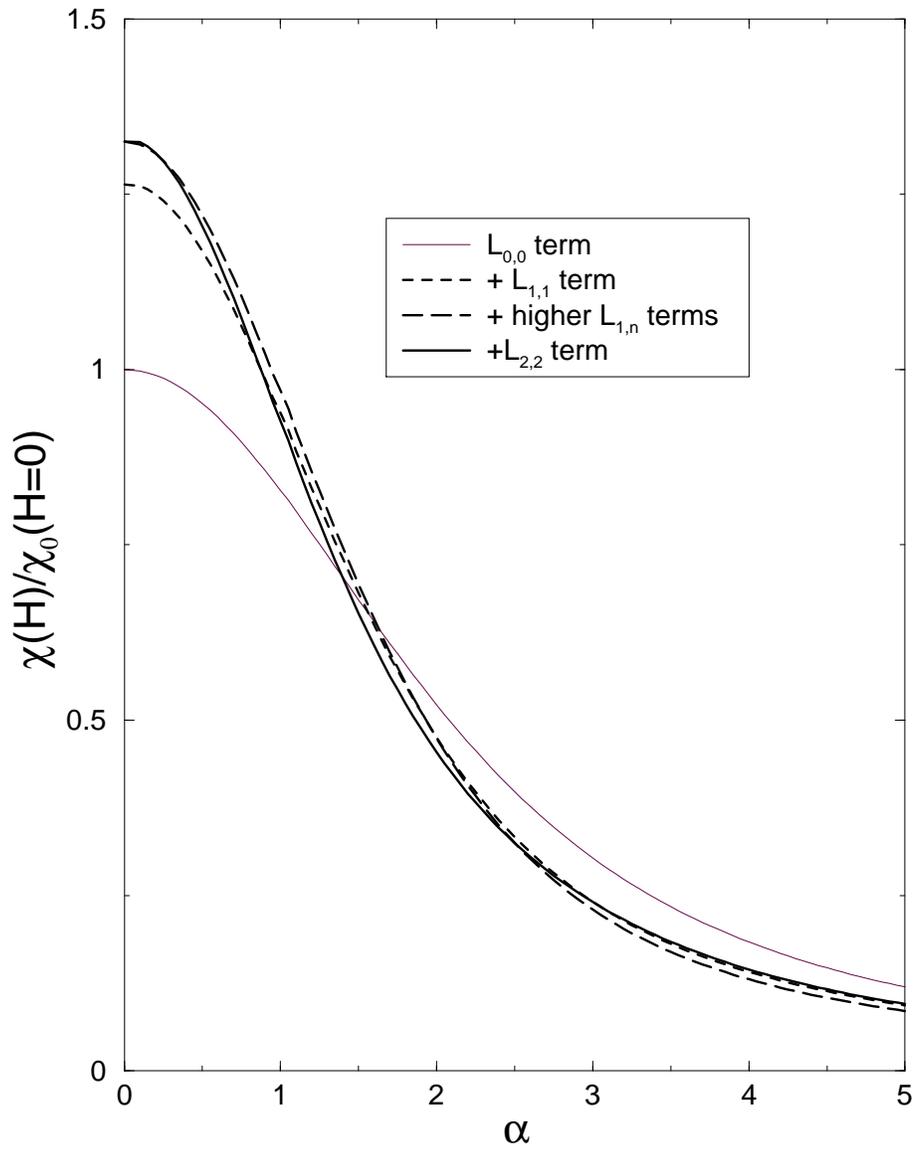
\begin{figure}\caption[]
{The reduced susceptibility for $\epsilon=2$ and
$\phi = 0.05$ as a function of $\alpha$. The higher order corrections are
largest at $\alpha = 0$. At moderate $\alpha$, the cancellation of the terms 
of order $\phi \epsilon^n$ with $n \geq 2$ against the term of order 
$\phi^2 \epsilon^2$ can again be seen.}
\label{suszOphi2}
\end{figure}
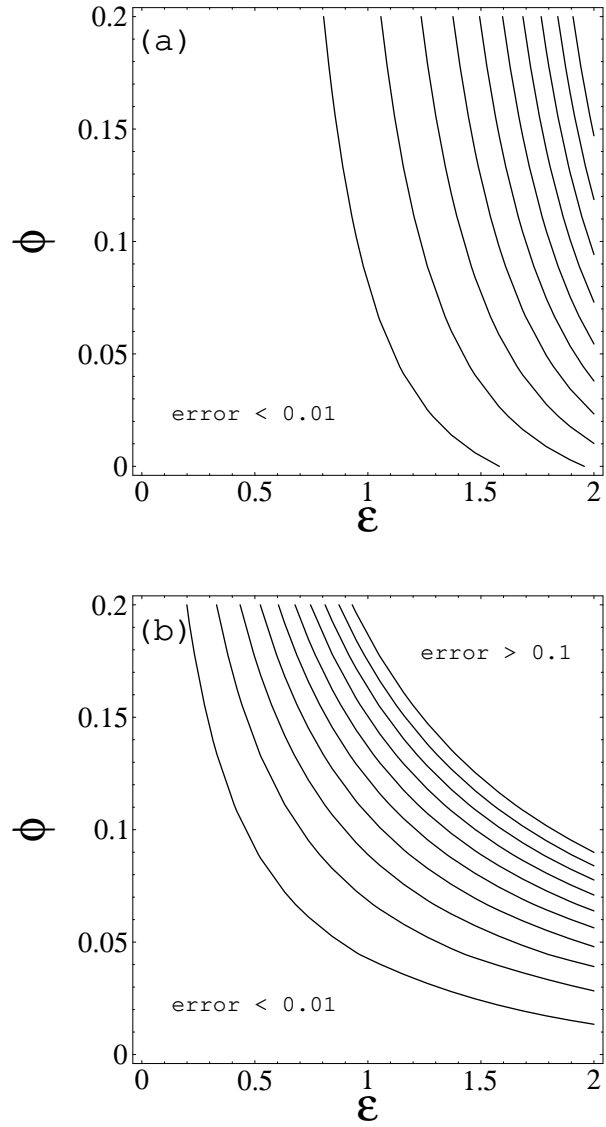
\begin{figure}\caption[]
{Quality of the lowest order expression (\ref{simpleM}) for the magnetization. 
(a) shows the isolines of the maximal -- with respect to $\alpha$ -- ratio
(\ref{epsnerror}) in steps of 0.01 and (b) shows those of the ratio 
(\ref{phi2error}).}
\label{errors}
\end{figure}
\begin{figure}\caption[]
{Comparison of the effects of polydispersity and of dipolar interaction. 
Plotted is the reduced magnetization versus $\alpha$ for different ferrofluid 
models:  a noninteracting monodisperse system (only $L_{0,0}$), a
noninteracting polydisperse system, and a monodisperse system with dipolar 
interaction for $\phi = 0.05$ and $\epsilon=2$.
The polydisperse system has a lognormal distribution of particle diameters 
(eq.~\ref{lognormal}) with 
a typical width of $\sigma = 0.3$ and the same third moment $\overline{D^3}$ as
the monodisperse fluid.}
\label{polymag}
\end{figure}

% \end{multicols}

\end{document}